\documentclass[aps, onecolumn, superscriptaddress, 12pt]{revtex4}
\usepackage{amssymb}
\usepackage{amsmath}
\usepackage{bm}
\usepackage[dvipdfmx]{graphicx}
\usepackage{dcolumn}
\usepackage{longtable}
\usepackage{color}
\usepackage[dvipdfm]{hyperref}
\def\vect#1{\mbox{\boldmath $#1$}}

\newcommand{\coznmn}{(Co$_{0.5}$Zn$_{0.5}$)$_{20-x}$Mn$_x$}
\newcommand{\ten}{Co$_{10}$Zn$_{10}$}
\newcommand{\nine}{Co$_9$Zn$_9$Mn$_2$}
\newcommand{\eight}{Co$_8$Zn$_8$Mn$_4$}
\newcommand{\seven}{Co$_7$Zn$_7$Mn$_6$}
\newcommand{\Tc}{$T_\mathrm{c}$}

\usepackage{ulem}

\begin{document}

\title{Metastable skyrmion lattices governed by magnetic disorder and anisotropy in $\beta$-Mn-type chiral magnets}
\author{K.~Karube}
\altaffiliation{These authors equally contributed to this work}
\affiliation{RIKEN Center for Emergent Matter Science (CEMS), Wako 351-0198, Japan.}
\author{J.~S.~White}
\altaffiliation{These authors equally contributed to this work}
\affiliation{Laboratory for Neutron Scattering and Imaging (LNS), Paul Scherrer Institute (PSI),
CH-5232 Villigen, Switzerland.}
\author{V.~Ukleev}
\affiliation{Laboratory for Neutron Scattering and Imaging (LNS), Paul Scherrer Institute (PSI),
CH-5232 Villigen, Switzerland.}
\author{C.~D.~Dewhurst}
\affiliation{Institut Laue-Langevin (ILL), 71 avenue des Martyrs, CS 20156, 38042 Grenoble cedex 9, France}
\author{R.~Cubitt}
\affiliation{Institut Laue-Langevin (ILL), 71 avenue des Martyrs, CS 20156, 38042 Grenoble cedex 9, France}
\author{A.~Kikkawa}
\affiliation{RIKEN Center for Emergent Matter Science (CEMS), Wako 351-0198, Japan.}
\author{Y.~Tokunaga}
\affiliation{Department of Advanced Materials Science, University of Tokyo, Kashiwa 277-8561, Japan.}
\author{H.~M.~R\o nnow}
\affiliation{Laboratory for Quantum Magnetism (LQM), Institute of Physics, \'Ecole Polytechnique F\'ed\'erale de Lausanne (EPFL), CH-1015 Lausanne,
Switzerland.}
\author{Y.~Tokura}
\affiliation{RIKEN Center for Emergent Matter Science (CEMS), Wako 351-0198, Japan.}
\affiliation{Tokyo College and Department of Applied Physics, University of Tokyo, Bunkyo-ku 113-8656, Japan.}
\author{Y.~Taguchi}
\affiliation{RIKEN Center for Emergent Matter Science (CEMS), Wako 351-0198, Japan.}

\begin{abstract}
\:

Magnetic skyrmions are vortex-like topological spin textures often observed in structurally chiral magnets with Dzyaloshinskii-Moriya interaction. 
Among them, Co-Zn-Mn alloys with a $\beta$-Mn-type chiral structure host skyrmions above room temperature.
In this system, it has recently been found that skyrmions persist over a wide temperature and magnetic field region as a long-lived metastable state, and that the skyrmion lattice transforms from a triangular lattice to a square one.
To obtain perspective on chiral magnetism in Co-Zn-Mn alloys and clarify how various properties related to the skyrmion vary with the composition, we performed systematic studies on \ten, \nine, \eight\ and \seven\ in terms of magnetic susceptibility and small-angle neutron scattering measurements.
The robust metastable skyrmions with extremely long lifetime are commonly observed in all the compounds.
On the other hand, preferred orientation of a helimagnetic propagation vector and its temperature dependence dramatically change upon varying the Mn concentration.
The robustness of the metastable skyrmions in these materials is attributed to topological nature of the skyrmions as affected by structural and magnetic disorder. Magnetocrystalline anisotropy as well as magnetic disorder due to the frustrated Mn spins play crucial roles in giving rise to the observed change in helical states and corresponding skyrmion lattice form.

\end{abstract}

\maketitle

\section{Introduction}
Non-collinear and non-coplanar spin textures have recently attracted much attention as a source of various emergent electromagnetic phenomena. 
Magnetic skyrmions, vortex-like spin textures characterized by an integer topological charge, are a prototypical example of such non-coplanar magnetic structures\cite{Bogdanov,Nagaosa,Muhlbauer,Yu_FeCoSi}, and are anticipated to be applied to spintronics devices since they can be treated as particles and driven by an ultra-low current density\cite{Jonietz,Yu_current,Iwasaki}.
Thus far, skyrmions have been observed in various magnets due to several microscopic mechanisms, such as competition between the Dzyaloshinskii-Moriya interaction (DMI) and ferromagnetic exchange interaction\cite{Muhlbauer,Yu_FeCoSi,Heinze,Romming,Matsuno,Woo,Soumyanarayanan,Yu_FeGe,Seki,Tokunaga,Li,Kezsmarki,Kurumaji_VOSe2O5,Nayak}, magnetic dipole interaction \cite{Yu_LSMO,Wang,Hou,Takagi_Cr11Ge19}, and magnetic frustration or Ruderman-Kittel-Kasuya-Yosida (RKKY) interaction\cite{Kakihana,Kurumaji_Gd2PdSi3,Max_Gd3Ru4Al12}.
Among them, the DMI arises from relativistic spin-orbit interaction and broken inversion symmetry either at interfaces of thin-film layers\cite{Heinze,Romming,Matsuno,Woo,Soumyanarayanan} or in bulk materials with noncetrosymmetric crystal structures\cite{Muhlbauer,Yu_FeCoSi,Yu_FeGe,Seki,Tokunaga,Li,Kezsmarki,Kurumaji_VOSe2O5,Nayak}. 

In the structurally chiral magnets as represented by B20-type compounds (e.g., MnSi\cite{Muhlbauer}, Fe$_{1-x}$Co$_x$Si\cite{Yu_FeCoSi}, FeGe\cite{Yu_FeGe}) and Cu$_2$OSeO$_3$\cite{Seki} with the space group of $P$2$_1$3, the DMI gradually twists ferromagnetically coupled moments to form a long-period helimagnetic state described by a magnetic propagation vector ($\vect{q}$ vector). 
The magnitude of $\vect{q}$ is given by $q$ $\propto$ $D/J$, where $D$ and $J$ correspond to the DMI constant and the exchange stiffness, respectively, and the propagation direction is determined by magnetic anisotropy.
Near the helimagnetic transition temperature $T_\mathrm{c}$, magnetic fields induce a triangular-lattice skyrmion crystal (SkX) as illustrated in Fig. 1(c), which is often described as a triple-$\vect{q}$ structure with the $\vect{q}$ vectors displaying mutual 120$^\circ$ angles perpendicular to the magnetic field. 
In general, SkX is stabilized by thermal fluctuations and thus its thermodynamical equilibrium state is confined to a narrow temperature and magnetic field region just below $T_\mathrm{c}$, and topologically-trivial helical or conical states are the thermodynamically most stable states at lower temperatures. 

Recently, Co-Zn-Mn alloys have been identified as a new class of chiral magnets based on bulk DMI which host skyrmions above room temperature\cite{Tokunaga}. 
The materials crystallize in a $\beta$-Mn-type chiral cubic structure with the space group of $P$4$_1$32 (defined as right-handed structure) or $P$4$_3$32 (left-handed structure), where 20 atoms per unit cell are distributed over two Wyckoff sites (8$c$ and 12$d$) as illustrated in Fig. 1(a). 
The 8$c$ sites are mainly occupied by Co atoms while the 12$d$ sites are mainly occupied by Zn and Mn\cite{Hori,Xie,Bocarsly,Nakajima_CoZnMn}.
The $\beta$-Mn-type structure forms in all the solid solutions of \coznmn\ from \ten\ to Mn$_{20}$ ($\beta$-Mn itself)\cite{Hori,Karube_776}. 

Magnetic phase diagram on the temperature ($T$) - Mn concentration ($x$) plane is reproduced from Ref. \cite{Karube_776} and displayed in Fig. 1(b) with additional information obtained in the present study.
One end member \ten\ shows a helimagnetic ground state with Ref. \cite{Tokunaga} reporting a magnetic periodicity $\lambda$ $\sim$ 185 nm below $T_\mathrm{c}$ $\sim$ 460 K. 
\Tc\ rapidly decreases as partial substitution of Mn proceeds, and \eight\ with \Tc\ $\sim$ 300 K exhibits a thermally equilibrium SkX state at room temperature under magnetic fields. 
The magnitude of the DMI constant in \eight\ has been experimentally evaluated to be $D$ $\sim$ 0.53 mJ/m$^2$, which is several times smaller than that in FeGe\cite{Takagi_DM}.
The DMI critically depends on band structure and electron band filling as demonstrated in Fe-doped \eight, where even a sign change in the DMI, namely, reversal of skyrmion helicity, occurs as the Fe concentration is increased\cite{Karube_Fe-dope}. 

Although the thermodynamical equilibrium SkX phase in \eight\ exists only in a narrow temperature and magnetic field region, it has been demonstrated that a once-created SkX can persist over the whole temperature region below room temperature and a wide magnetic field region as a long-lived metastable state via a conventional (a few K/min) field cooling (FC)\cite{Karube_884}. Moreover, the lattice form of the metastable SkX undergoes a reversible transition, accompanied by large increase in $q$, from a conventional triangular lattice to a novel square one, which is described as an orthogonal double-\vect{q} state, as illustrated in Fig. 1(e)\cite{S-SkX}. 
Similar robust metastable SkX has been observed in \nine\ with $T_\mathrm{c}$ $\sim$ 400 K, where the metastable SkX persists even at zero field above room temperature\cite{Karube_992}.
By means of Lorentz transmission electron microscopy (LTEM) for thin-plate specimens, various exotic skyrmion-related structures, such as I- or L-shaped elongated skyrmions\cite{Morikawa} [Fig. 1(g)], a smectic liquid-crystalline structure of skyrmions\cite{Nagase} and a meron-antimeron square lattice\cite{Yu_893}, have been observed.

The other end member $\beta$-Mn is well known as a spin liquid, while the lightly doped $\beta$-Mn alloys with slight disorder exhibit a spin glass, due to geometrical frustration among antiferromagnetically coupled Mn spins in the hyper-kagome network of the 12$d$ sites\cite{Nakamura,Stewart_betaMn,Stewart_betaMnAl,Stewart_betaMnCo,Stewart_betaMnIn,Paddison}. 
Therefore, \coznmn\ possesses both magnetic frustrations inherent to $\beta$-Mn and magnetic disorder due to the mixture of ferromagnetic Co spins and antiferromagnetic Mn spins, which give rise to a spin glass phase below $T_\mathrm{g}$ over a wide range of the Mn concentration (3 $\leq$ $x$ $\leq$ 19)\cite{Karube_776}.
In particular, the spin glass phase invades the helical phase for 3 $\leq$ $x$ $\leq$ 7 displaying reentrant spin glass behavior\cite{Note_RSG} as similarly reported for a number of ferromagnets\cite{Coles,Motoya,Maletta,Hanasaki,Ronnow,Mirebeau}, antiferromagnets\cite{Yoshizawa,Leavey} and helimagnets\cite{Sato} with chemical and magnetic disorder. 
The spin glass nature was confirmed by previous frequency-dependent ac susceptibility measurements for \seven\ ($T_\mathrm{c}$ $\sim$ 160 K, $T_\mathrm{g}$ $\sim$ 30 K) \cite{Footnote_776,Karube_776,Bocarsly}.
It has been discovered that a novel equilibrium phase of disordered skyrmions exists just above $T_\mathrm{g}$ in \seven, which is thermodynamically disconnected from the conventional equilibrium SkX phase just below $T_\mathrm{c}$, and presumably stabilized by a cooperative interplay between the chiral magnetism with DMI and the frustrated magnetism\cite{Karube_776}.

Despite these extensive studies for Co-Zn-Mn alloys, it remains elusive how the (meta)stability and lattice form of skyrmions vary with the Mn concentration from the end member \ten, and how the two magnetic elements of Co and Mn contribute to the skyrmion formation and the transformation of skyrmion lattice.
In order to obtain a more complete perspective on the skyrmion states in the Co-Zn-Mn alloys, 
here we present new data sets obtained by small-angle neutron scattering (SANS) and magnetic susceptibility that have not been reported previously. From the new data we report: (i) Identification of the equilibrium skyrmion phase, and temperature and field dependence of helical and metastable skyrmion states in \ten, (ii) an off-axis magnetic field experiment for metastable skyrmions in \eight\ to reveal the role of anisotropy, (iii) temperature and field dependence of metastable skyrmions in \nine\ below ambient temperature down to the lowest temperature to clarify the effect of dilute Mn moments that becomes significant only at low temperatures, (iv) temperature and field evolution of heavily disordered metastable skyrmions in \seven, and (v) lifetime of metastable skyrmions in \ten\ and \eight. Taking the new results together with our previous ones, we show systematic changes in lattice forms and lifetimes of the metastable skyrmions, as well as the key roles of magnetic disorder and anisotropy, as functions of temperature and Mn concentration.
As summarized in Fig. 2, the metastable skyrmion state prevails in a wide temperature and field range for all the materials.
In \ten, triangular lattice of skyrmions transforms to rhombic-like [Fig. 1(d)] as the temperature is lowered in a low field. 
This triangular-rhombic structural transformation is governed by enhanced magnetocrystalline anisotropy that favors $\vect{q}$ $\parallel$ $<$111$>$. 
On the other hand, in the Mn-doped compounds, $\vect{q}$-vector orientation changes from $<$111$>$ to $<$100$>$, and structural transformation from triangular lattice to square one occurs at low temperatures and low fields. As the Mn concentration is increased, and as the helimagnetic \Tc\ thus falls, the antiferromagnetic correlations of Mn spins start to develop on cooling from higher temperatures. The resulting magnetic disorder drives a large increase in $q$ value, thereby triggering the transformation to the square lattice state, in cooperation with the enhanced magnetic anisotropy toward $\vect{q}$ $\parallel$ $<$100$>$ at low temperatures.

The format of this paper is as follows. After we describe experimental methods in Section II, results and discussion are presented in Section III, and conclusion is given in Section IV.
In the section III, we first overview helical and skyrmion states in all the compounds (Subsections A, Figures 2, 3, Table 1).
Then, we show detailed results of magnetic susceptibility and SANS measurements for each composition: \ten\ (Subsection B, Figures 4-6), \eight\ (Subsection C, Figure 7), \nine\ (Subsection D, Figure 8), and \seven\ (Subsection E, Figure 9-11), and summarize all of them (Subsection F).
Finally, we present composition dependence of lifetime of metastable skyrmions (Subsection G, Figures 12, 13).
With all these results, we discuss the roles of magnetic disorder and anisotropy in leading to the robust metastable skyrmions and their novel lattice forms in Co-Zn-Mn alloys.

\section{Experimental methods}
\subsection{Sample preparation}

Single-crystalline \ten\ was grown by a self-flux method in an evacuated quartz tube.
The single crystals were cut along the (110), (-110) and (001) planes with a rectangular shape for magnetization and ac susceptibility measurements as well as SANS measurement after the crystalline orientation was determined by the X-ray Laue diffraction method. 
Single crystals of \nine, \eight\ and \seven\ were grown by the Bridgman method as described in our previous papers\cite{Karube_884,Karube_992,Karube_776}, and were cut along the (110), (-110) and (001) planes for \nine, and along the (100), (010) and (001) planes for \eight\ and \seven, respectively.

\subsection{Magnetization and ac susceptibility measurements}

Magnetization and ac susceptibility measurements were performed with a superconducting quantum interference device magnetometer (MPMS3, Quantum Design). 
High-temperature measurements above 400 K for \ten\ were performed by using an oven option.
In the ac susceptibility measurements, the ac drive field was set as $H_{ac}$ = 1 Oe for all the compounds, and the ac frequency was selected to be $f$ = 17 Hz for \ten\ and $f$ =193 Hz for \nine, \eight\ and \seven.
Magnetic fields were applied along the [110] direction for \ten\ and \nine, and along the [100] direction for \eight\ and \seven, respectively.
Due to the difference in the shape between the samples used in the ac susceptibility (field parallel to the plate) and the SANS measurements (field perpendicular to the plate), their demagnetization factors are different. 
To correct for this difference so that a common absolute magnetic field scale is used throughout this paper, the field values for the ac susceptibility measurements are calibrated as $H_\mathrm{c} = NH$ where $N$ = 3.0, 2.0, 3.7 and 2.7 for \ten, \nine, \eight\ and \seven, respectively.

\subsection{Small-angle neutron scattering (SANS) measurements}

SANS measurements for \ten, \nine\ and \eight\ were performed using the SANS-I instrument at the Paul Scherrer Institute (PSI), Switzerland.
For high-temperature measurements above 300 K for \ten\ and \nine, a bespoke oven stick designed for SANS experiments was used. 
SANS measurements of \seven\ were done using the D33 instrument at the Institut Laue-Langevin (ILL), France. 
The neutron wavelength was selected to be 10 $\mathrm{\AA}$ with a 10\% full width at half maximum (FWHM) spread in all the measurements.
For all the SANS data shown here, nuclear and instrumental background signals have been subtracted by using data taken either well above the helimagnetic ordering temperature \Tc, or well above the polarized ferromagnetic transition field.

For \ten, the mounted single-crystalline sample was installed into a horizontal field cryomagnet so that the incident neutron beam ($k_\mathrm{i}$) and the magnetic field ($H$) were parallel to the [110] direction.
The cryomagnet was rotated (`rocked') together with the sample around the vertical [001] direction, and the rocking angle ($\omega$) between $k_\mathrm{i}$ and $H$ was scanned from $-$20$^\circ$ to 20$^\circ$ by 2$^\circ$ step. Here, $\omega$ = 0$^\circ$ corresponds to the $k_\mathrm{i}$ $\parallel$ $H$ configuration. 
The observed FWHM of the rocking curves (scattering intensity versus $\omega$) was always broader than 19$^\circ$.
Therefore, to deduce accurately the relative intensities and positions of all of the Bragg spots in single images, all the SANS images displayed in this paper are obtained by summing multiple SANS measurements taken over $-$8$^\circ$ $\leq$ $\omega$ $\leq$ 8$^\circ$. 

As detailed in our previous papers\cite{Karube_992,Karube_884,Karube_776}, similar SANS measurements were performed with the $k_\mathrm{i}$ $\parallel$ $H$ $\parallel$ [110] configuration for \nine, and $k_\mathrm{i}$ $\parallel$ $H$ $\parallel$ [001] configuration for \eight\ and \seven. 

\section{Results and Discussion}
\subsection{Overview of state diagrams}
First, we briefly overview the results of basic magnetic properties in helimagnetic states and metastable skyrmion states for all the compounds. 

\subsubsection{Helical state}
Figure 3(a-d) shows temperature ($T$) dependence of magnetization ($M$) under a small magnetic field of 20 Oe. 
In \ten, $M$ shows a sharp increase due to a helimagnetic transition at $T_\mathrm{c}$ $\sim$ 414 K, and then stays almost independent of temperature both for the field cooling (FC) and the zero-field-cooled field-warming (ZFC-FW) processes.
The other compounds exhibit gradual decrease in $M$ upon cooling at some temperature region ($T_\mathrm{L}$ $\leq$ $T$ $\leq$ $T_\mathrm{H}$).
\nine\ shows almost temperature-independent $M$ in a wide temperature range below $T_\mathrm{c}$ $\sim$ 396 K, while the gradual decrease is observed below $T_\mathrm{H}$ $\sim$ 50 K. 
In \eight, $T$-independent behavior below $T_\mathrm{c}$ $\sim$ 299 K is followed by the gradual decrease in $M$ upon cooling from $T_\mathrm{H}$ $\sim$ 120 K down to $T_\mathrm{L}$ $\sim$ 40 K. 
A sharp drop of $M$ at $T_\mathrm{g}$ $\sim$ 9 K observed only in the ZFC-FW process is due to a reentrant spin-glass transition. 
\seven\ exhibits the gradual decrease in $M$ below $T_\mathrm{H}$ $\sim$ 120 K down to $T_\mathrm{L}$ $\sim$ 50 K below the helimagnetic transition at $T_\mathrm{c}$ $\sim$ 158 K. The reentrant spin glass transition is observed around $T_\mathrm{g}$ $\sim$ 26 K as manifested in a large deviation between FC and ZFC-FW processes. 
The reentrant spin glass transition temperature $T_\mathrm{g}$ determined by the $M(T)$ curve is very close to the zero frequency limit of $T_\mathrm{g}$ determined by ac susceptibility curves with various frequencies as previously reported\cite{Footnote_776}. $T_\mathrm{g}$ does not show significant dependence on the magnetic field below the field-induced ferromagnetic region as plotted in the phase diagrams in Fig. 2(d, e).
These characteristic temperatures ($T_\mathrm{c}$, $T_\mathrm{H}$, $T_\mathrm{L}$ and $T_\mathrm{g}$) are plotted in the $T$-$x$ phase diagram in Fig. 1(b). 

Magnetization curves at 2 K that characterize the ground states are shown in Supplementary Fig. S1\cite{Supple}.
The values of \Tc\ and saturation magnetization $M_\mathrm{s}$ at 2 K and at 7 T in all the compounds are summarized in Table 1.
From the value of $M_\mathrm{s}$ $\sim$ 11.3 $\mu_\mathrm{B}$/f.u. in \ten, magnetic moment at Co site is estimated to be $\sim$ 1.1 $\mu_\mathrm{B}$.
While \Tc\ monotonically decreases with increasing Mn concentration, $M_\mathrm{s}$ displays a maximum in \nine. 

The characteristic $T$-dependence of $M$ at 20 Oe is compared with that of the magnitude of helical wavevector ($q$) from SANS measurements in Fig. 3(e-h).
It turns out that the gradual decrease in $M$ from $T_\mathrm{H}$ to $T_\mathrm{L}$ corresponds to the large increase in the value of $q$, i.e. the decrease in the helimagnetic periodicity $\lambda$ (= $2\pi/q$).
The largest value of $\lambda$ at high temperatures ($\lambda_\mathrm{max}$) and the smallest value of $\lambda$ at low temperatures ($\lambda_\mathrm{min}$) for each compound are summarized in Table 1. 
We also present the preferred orientation of $\vect{q}$ in Table 1. 
While the preferred $\vect{q}$ direction is $<$111$>$ for \ten, it is parallel to $<$100$>$ in \nine, \eight\ and \seven, as detailed in the following sections. 

\subsubsection{Metastable skyrmion state}

Figure 2 shows the $T$-$H$ diagrams determined by ac susceptibility and SANS measurements for each composition.
In these diagrams, the equilibrium SkX phase and the metastable SkX state are presented together, as determined by field scans after ZFC and by field scans after a FC via the equilibrium phase, respectively, as schematically illustrated in Fig. 2(a).
The robust metastable skyrmion state that exists over a very wide temperature and field region is commonly observed from \ten\ to \seven. 
In \ten\, the triangular lattice of metastable skyrmions (M-T-SkX) distorts and transforms to a rhombic one (M-R-SkX) at low fields during the FC.
On the other hand, the lattice form changes to a square one (M-S-SkX) at low temperatures and low fields in the other Mn-doped compounds. 
As the partial Mn substitution proceeds, the phase space of the metastable triangular SkX state (M-T-SkX) is squeezed, accompanied by the suppression of $T_\mathrm{c}$, while the region occupied by square SkX state (M-S-SkX) expands toward higher temperatures.
The transition to the square SkX is accompanied by a large increase in $q$ as similarly observed in the helical states [Fig. 3(f-h)]. 

\subsection{\ten}
In this section, we present detailed results of SANS and ac susceptibility measurements for \ten\ and show that the preferred orientation of $\vect{q}$ vector is $<$111$>$ direction, and that on field cooling metastable skyrmion lattice distorts and undergoes a structural transformation from a conventional triangular lattice to a rhombic one.

\subsubsection{Identification of equilibrium SkX}
First, we identify an equilibrium SkX phase (Fig. 4). 
Photographs of single-crystalline samples used for SANS and ac susceptibility measurements are shown in Fig. 4(a).
Figure 4(b) shows the field variation of SANS images at 410 K.
The scattering signal observed at 0 T is attributed to a helical multi-domain state.
Here, 4 broad spots are observed but the peak positions are not aligned clearly to any unique set of high-symmetry crystal axes. 
This peculiar scattering distribution near \Tc\ may be attributed to rather flexible nature of $\vect{q}$ vector due to the negligible anisotropy at high temperatures under an influence of possible residual local strain of the sample.
Under magnetic fields parallel to the incident neutron beam, the SANS signal diminishes at 0.02 T most likely due to the transition to a conical state whose $\vect{q}$ vector is parallel to the field and thus not detected in this configuration. 
At 0.03 T, a broad 6-spot pattern, a hallmark of triangular lattice of skyrmions (triple-$\vect{q}$ $\perp$ $H$), appears. 
As the magnetic field is further increased to 0.06 T, the conical state is stabilized again and the signal disappears.
SANS intensity perpendicular to the field is plotted against field in Fig. 4(c).
The SANS intensity exhibits a peak at 0.03 T, where the volume fraction of skyrmions is maximized.

The field dependence of the real- ($\chi^\prime$) and imaginary-part ($\chi^{\prime\prime}$) of the ac susceptibility at 410 K is shown in Fig. 4(d).
A dip structure is observed in $\chi^\prime$ between 0.02 T and 0.05 T where in addition $\chi^{\prime\prime}$ exhibits a clear double peak structure. These features correspond to a SkX state surrounded by a conical state, and show a good agreement with the region of the enhanced SANS intensity.
Therefore, the phase boundaries of SkX state are determined as the peak positions at the both sides of the dip structure in $\chi^\prime$.
Contour plots of $\chi^\prime$ and $\chi^{\prime\prime}$ with the phase boundaries on the $T$-$H$ plane are presented in Fig. 4(e) and (f), respectively. 
The equilibrium SkX phase exists in a narrow temperature region from $T_\mathrm{c}$ $\sim$ 412 K down to 407 K, below which the field-induced conical state is the most stable.

\subsubsection{Temperature evolution of helical state in zero field}
Next, we discuss temperature variation of the helical state (Fig. 5).
Figure 5(b) shows the SANS patterns at selected temperatures on zero-field cooling from 410 K to 1.5 K. 
As temperature is lowered from \Tc\ down to 300 K, the SANS intensity gradually accumulates along the [1-11] direction due to enhanced magnetic anisotropy. 
Upon further decreasing temperature down to 1.5 K, the SANS intensity splits into 4 spots: 2 spots with stronger intensity along the [1-11] direction, and the other 2 spots with weaker intensity along the [-111] direction. 
This 4 spot pattern at 1.5 K is expected for the multi-domain helical state with $\vect{q}$ $\parallel$ $<$111$>$ as schematically illustrated in Fig. 5(a). 
Notably, the preferred orientation of $\vect{q}$ vector in \ten\ is different from those of the other \coznmn\ with $x$ $\geq$ 2, which is found to be $<$100$>$ by our previous studies\cite{Karube_884,Karube_992,Karube_776}.

Figure 5(c) shows the temperature dependence of the SANS intensity integrated for the [-111] and [1-11] directions ($\phi$ = 55$^\circ$, 125$^\circ$, 235$^\circ$ and 305$^\circ$), where conventional order-parameter like evolution of the intensity for $\vect{q}$ $\parallel$ $<$111$>$ is clearly observed. 
The radial $q$ dependence of the SANS intensity for the direction close to [1-11] is displayed in Fig. 5(d). 
The peak center slightly shifts toward higher $q$ region and the peak width slightly increases on cooling.
The helical $q$ value, which is determined as the peak center of a Gaussian function fitted to the intensity vs $q$ curve [solid line in Fig. 5(d)], and its FWHM are plotted against temperature in Fig. 3(e) and (i), respectively.
Both the $q$ value and the FWHM gradually increases on cooling but the variation is much smaller than that in the Mn-doped compounds.
The helical periodicity estimated as $\lambda = 2\pi/q$ varies a little from 156 nm to 143 nm on cooling as summarized in Table 1.
Note that the values of \Tc\ = 414 K and $\lambda$ = 156 nm for a single crystal of \ten\ in the present study are smaller than the values (\Tc\ = 462 K and $\lambda$ = 185 nm) for a polycrystalline sample in the previous study\cite{Tokunaga}, probably due to a slight difference in the composition. 

As detailed in Supplementary Fig. S4\cite{Supple}, we also find that the helical state at 1.5 K forms a chiral soliton lattice\cite{Togawa} under magnetic fields due to the enhanced magnetocrystalline anisotropy which enforces the $\vect{q}$ vector along $<$111$>$, and perpendicular to the field.

\subsubsection{Temperature dependence of metastable SkX}
Next, we discuss temperature variation of the metastable SkX state (Fig. 6).
The SANS patterns in a FC at 0.03 T from 412 K to 1.5 K are displayed in Fig. 6(b). 
In order to gain the sufficient cooling rate across the boundary between the equilibrium SkX and conical phases, the magnetic field was applied at 412 K (slightly higher than presented in Fig. 4), where the observed SANS pattern is ring-like, indicating that the triangular lattice of skyrmions is orientationally disordered.
As the temperature is lowered, the ring-like SANS pattern gradually changes to broad 4 spots along the [-111] and [1-11] directions.
The scattering intensity is plotted as a function of azimuthal angle $\phi$ in Fig. 6(c).
Clearly, 4 peaks around $\phi$ = 55$^\circ$, 125$^\circ$, 235$^\circ$ and 305$^\circ$ emerge out of a rather featureless profile at 412 K as the temperature is lowered down to 1.5 K\cite{Footnote_1010}.

The 4-spot SANS pattern observed at low temperatures is expected for a rhombic lattice of skyrmions with double-$\vect{q}$ vectors $\parallel$ $<$111$>$ that are rotated by 110$^\circ$ from each other as schematically illustrated in Fig. 6(a) and Fig. 1(d). 
The SANS intensity integrated over the region close to the [-111] and [1-11] directions under the FC is plotted against temperature together with the intensity integrated around the [001] and [-110] directions in Fig. 6(d).
Below 360 K the intensity around the [-111] and [1-11] directions becomes higher than that around the [001] and [-110] directions. This temperature corresponds to the onset of triangular-rhombic lattice structural transition and is plotted with a purple diamond (right one) in the state diagram in Fig. 2(b). 
The broad SANS pattern and the gradual temperature evolution indicate coexistence of the triangular SkX state and the rhombic one over a wide temperature region.

The transformation to the rhombic lattice is probably driven by significant enhancement of magnetocrystalline anisotropy at low temperatures that favors $\vect{q}$ $\parallel$ $<$111$>$ as found in the helical state (Fig. 5).
It is also noted that the absolute value of $q$ slightly increases upon lowering temperature as plotted in Fig. 3(e).
The relation between the lattice transformation and the change in the $q$ value is further discussed in the summary section of metastable SkX (Subsection F-2).

\subsubsection{Metastability of skyrmions against field variation}
We also investigated the metastability and the lattice form of skyrmions against field variation.
The detailed field-dependent SANS data at 300 K after the FC (0.03 T) is presented in Supplementary Fig. S2\cite{Supple}. 
In the field sweepings toward positive direction, the broad 4-spot pattern with stronger intensity at the [-111] and [1-11] directions changes to a uniform ring above 0.1 T as observed in the equilibrium SkX phase at 412 K and 0.03 T. 
This field variation corresponds to the change in skyrmion lattice form from the rhombic one to the original disordered triangular one. 
Importantly, such a ring-like pattern is never observed upon the field sweeping at 300 K after ZFC.
This result ensures that the broad 4-spot pattern appearing after the FC is attributed to the metastable rhombic SkX (M-R-SkX) and distinct from helical multi-domain state.
The SANS intensity originating from the disordered triangular SkX persists up to the region close to the field-induced ferromagnetic phase in contrast to the helical state. 
In the field sweeping toward negative direction, on the other hand, the scattering intensity around the [-111] and [1-11] directions is further increased and the 4-spot pattern becomes clearer. This indicates that the rhombic lattice is more stable than the triangular one at zero or negative fields. As the field is swept further negative, finally the SANS intensity from the rhombic SkX disappears upon the transition to the equilibrium conical state.

The field variation at 1.5 K after the FC (0.03 T) is displayed in Supplementary Fig. S3\cite{Supple}, where similar change in the SANS pattern was observed while the 4 broad spots persist over a wider field region than that observed at 300 K.

The field variations of SANS at 300 K and 1.5 K are consistent with those of ac susceptibility, where smaller values (as compared with conical state) corresponding to the metastable SkX are observed, accompanied by a characteristic asymmetric hysteresis.
Similar field-dependent ac susceptibility measurements after FC processes were performed at different temperatures, and boundaries of the metastable SkX state are plotted in the $T$-$H$ phase diagram in Fig. 2(b).
The metastable SkX state persists over a wide $T$-$H$ region, including room temperature and zero field.
Within the metastable state, the lattice form of skyrmions undergoes the transition from triangular to rhombic at a low-$T$ and low-$H$ region where $\vect{q}$ vector anisotropy along $<$111$>$ is significant. 

\subsection{\eight}
In this section, we describe detailed results of SANS measurements with off-axis field configuration in \eight\ and provide evidence for increased magnetocrystalline anisotropy that favors $\vect{q}$ $\parallel$ $<$100$>$ at low temperatures.

\subsubsection{Temperature-driven structural transition of metastable skyrmion lattice}
First, we review the temperature variation in metastable SkX under a field parallel to the [001] direction, as reported in our previous paper\cite{Karube_884} (see Supplementary Fig. S5 for the details\cite{Supple}).
In a FC process at 0.04 T, SANS pattern transforms from 12 spots to 4 spots below 120 K.
The 12-spot pattern at high temperatures corresponds to a 2-domain triangular SkX state, in which one of triple-$\vect{q}$ is parallel to the [010] or [100] direction. 
The 4-spot pattern at low temperatures is attributed to a square SkX state with double-$\vect{q}$ vectors parallel to the [010] and [100] directions. 
This temperature variation is reversible and the triangular SkX revives already at 200 K in the subsequent re-warming process, which rules out the possibility of relaxation to a helical multi-domain state. 
It is noted that the transformation to the square SkX below 120 K is accompanied by a large increase in $q$ as presented in Fig. 3(g).

\subsubsection{Temperature-dependent magnetic anisotropy as revealed by an off-axis field measurement}
Next, we show how $\vect{q}$ vector orientation changes during the triangular-square structural transition of skyrmion lattice under an off-axis magnetic field (Fig. 7).
The experimental configuration for the SANS measurement is illustrated in Fig. 7(a).
The magnetic field of 0.03 T was applied along the direction tilted away from the [001] direction by 15$^\circ$. 
While keeping this configuration, the cryomagnet and the sample were rotated together around the vertical [010] direction.
Here, $\omega$ is defined as a rocking angle between the incident neutron beam ($k_\mathrm{i}$) and the applied magnetic field ($H$), namely $\omega$ = 0$^\circ$ for $k_\mathrm{i}$ $\parallel$ $H$ and $\omega$ = $-$15$^\circ$ for $k_\mathrm{i}$ $\parallel$ [001].

Rocking curves (SANS intensity versus $\omega$) at selected temperatures are presented in Fig. 7(b).
Here, integrated intensity at $\phi$ = 90$^\circ$ and 270$^\circ$ is plotted against $\omega$.
For the equilibrium triangular SkX state, a rocking curve is expected to take a maximum at $\omega$ = 0$^\circ$ because triple-$\vect{q}$ are usually perpendicular to the applied field regardless of the crystal orientation.
However, the observed rocking curve at 295 K [red symbols in Fig. 7(b)] shows a broad maximum around $\omega$ $\sim$ 20$^\circ$. This indicates that the effective magnetic field inside the sample ($H_\mathrm{eff}$) is further tilted from the external field to the opposite side of the [001] axis as shown in Fig. 7(a), which can be understood in terms of a demagnetization effect in the present rectangular-shaped sample\cite{Demag}. 
As the temperature is lowered, the peak position of the rocking curve shifts to lower angle and eventually locates at $-$15$^\circ$ at 40 K. 

Temperature variations of the SANS patterns for $k_\mathrm{i}$ $\parallel$ $H_\mathrm{eff}$ and $k_\mathrm{i}$ $\parallel$ [001], averaged over the rocking angles of 14$^\circ$ $\leq$ $\omega$ $\leq$ 20$^\circ$ and $-$20$^\circ$ $\leq$ $\omega$ $\leq$ $-$10$^\circ$, respectively, are shown in Fig. 7(c) and (d).
For $k_\mathrm{i}$ $\parallel$ $H_\mathrm{eff}$, a 6-spot pattern is observed at 295 K. 
This 6-spot pattern corresponds to a single-domain triangular SkX state, in which triple-$\vect{q}$ are perpendicular to $H_\mathrm{eff}$ and one of them is parallel to the vertical [010] direction. 
The 6 spots are still discerned at 200 K as the triangular SkX persists as a metastable state.
At 120 K, the side 4 spots become much weaker as compared with the vertical 2 spots, and finally the 6-spot pattern is not discerned at 40 K.
For $k_\mathrm{i}$ $\parallel$ [001], on the other hand, only 2 vertical spots out of the 6 spots are observed at 295 K. 
Below 120 K, intensities around the horizontal region increase, and finally a 4-spot pattern is observed at 40 K.

The SANS intensity for the two different configurations is plotted against temperature in Fig. 7(e), clearly showing the change in intensity distribution below 120 K from the side 4 spots perpendicular to $H_\mathrm{eff}$ (blue circles) to the side 2 spots parallel to the [100] direction (red squares), in good accord with the large shift of the peak position in the rocking curves presented in Fig. 7(b).
Therefore, the double-$\vect{q}$ in the square SkX at low temperatures are aligned to the [010] and [100] direction, and the latter is not perpendicular to $H_\mathrm{eff}$.
This is in marked contrast to the triangular SkX state at high temperatures where all the triple-$\vect{q}$ are perpendicular to $H_\mathrm{eff}$.
Thus, the triangular-square SkX transition under the off-axis field is accompanied by a reorientation of skyrmions as schematically illustrated in Fig. 7(f).
This result indicates that the magnetic anisotropy favoring $\vect{q}$ $\parallel$ $<$100$>$ is strongly enhanced during the transformation to the square SkX as the temperature is reduced.

\subsection{\nine}
In this section, we present how the $\vect{q}$ vector evolves in terms of magnitude and orientation as a function of temperature in \nine.
In brief, the change in $\vect{q}$ direction, and the variation of $q$ value, are qualitatively similar to those found in \eight\ but take place at lower temperatures.

\subsubsection{Temperature dependence of metastable SkX}
First, we show temperature variation of the metastable SkX state (Fig. 8).
Selected SANS patterns in a FC process at 0.04 T from 390 K to 10 K are displayed in Fig. 8(b). 
In this measurement, the magnetic field and the incident neutron beam are parallel to the [110] direction.
At 390 K within the equilibrium SkX phase, the SANS pattern shows 6 spots, corresponding to a triangular SkX with one of triple-$\vect{q}$ $\parallel$ [001] as illustrated in the right panel of Fig. 8(a). 
The triangular SkX persists down to 100 K as a metastable state during the FC.
Below 50 K, the signal from the side 4 spots becomes weaker as compared with the vertical 2 spots, as similarly observed in the off-axis field measurement for \eight\ (Fig. 7(b)).

Figure 8(d) presents the temperature dependence of the SANS intensity for the side 4 spots and the vertical 2 spots, clearly showing the cross-correlation between the reduced intensity for the side 4 spots and the increased intensity for the [001] direction below 50 K.
Rocking curves during the FC are displayed in Fig. 8(c).
While the rocking curve exhibits a peak around $\omega$ = 0$^\circ$ above 100 K, the intensity around $\omega$ = 0$^\circ$ decreases below 50 K and finally the rocking curve forms a concave shape around $\omega$ = 0$^\circ$ at 10 K.
These results are similar to those observed in the SANS pattern for \eight\ under the off-axis field as shown in Fig. 7.
Therefore, we attribute the change in the SANS pattern below 50 K in \nine\ to the $\vect{q}$ vector reorientation and associated transition from a triangular SkX to a square SkX.
Namely, one of the double-$\vect{q}$ in the square SkX is aligned to the [001] direction but the other is aligned to [100] or [010] directions that are out of the (110) plane, resulting in 2 vertical spots on the (110) plane as illustrated in the left panel of Fig. 8(a).

The radial $q$ dependence of the SANS intensity for the [001] direction is shown in Fig. 8(e). 
While the peak center ($q$ $\sim$ 0.05 nm$^{-1}$) is almost temperature independent above 100 K, a large shift up to $q$ $\sim$ 0.07 nm$^{-1}$ is observed below 50 K, accompanied by a slight broadening of the width. 
Thus, the transition from the triangular SkX to the square SkX in \nine\ is also accompanied by the increase in the magnitude and width of the $\vect{q}$ vector, again similarly as observed in \eight.

\subsubsection{Field dependence of metastable SkX}

Next, we discuss field variation in the $\vect{q}$ vector in the metastable SkX state at 10 K after the FC under 0.04 T (see Supplementary Fig. S6 for the details\cite{Supple}).
With increasing field from 0.04 T to 0.3 T, the 2-spot SANS pattern gradually changes to a ring-like one, in which the scattering intensity lies in the (110) plane as similarly observed at high temperatures.
This indicates that the square SkX with $\vect{q}$ $\parallel$ $<$100$>$ at low temperatures transforms to an orientationally disordered triangular SkX with $\vect{q}$ $\perp$ $H$ at high fields.
 
Taking the above results into account, we summarize the state diagram of the metastable SkX in \nine\ in Fig. 2(c), which is characterized by a large region of the triangular SkX state and a small region of the square SkX state existing only at low temperatures below $\sim$ 50 K and low fields.

\subsection{\seven}
In this section, we discuss how the magnetic state is affected by the increased Mn concentration in \seven.

\subsubsection{Heavily-disordered square lattice within the metastable state}
First, we discuss the temperature variation of the metastable SkX (Fig. 9).
Figure 9(b) shows the change in the SANS patterns during a FC at 0.025 T from 146 K to 1.5 K. 
In this measurement, the magnetic field and the incident neutron beam are parallel to the [001] direction.
At 146 K within the equilibrium SkX phase, a 12-spot pattern corresponding to a 2-domain triangular SkX with one of triple-$\vect{q}$ $\parallel$ [010] or [100], as schematically illustrated at right panel in Fig. 9(a), is observed. 
However, the spots display a significant azimuthal broadening as compared with those observed from the equilibrium SkX in \eight\ sample. 
The 12-spot pattern changes to a pattern of 4 broadened spots at 100 K, which corresponds to a metastable square SkX with double-$\vect{q}$ $\parallel$ [010] and [100] as shown at left panel in Fig. 9(a). 
With further decreasing temperature down to 60 K, the 4 spots become even broader. 
This pattern of 4 very broad spots remains almost unchanged across the reentrant spin glass transition ($T_\mathrm{g}$ $\sim$ 30 K) down to 1.5 K.

The SANS intensity from the metastable SkX is plotted as a function of temperature in Fig. 9(c). 
Above 130 K, the intensities integrated for directions close to $<$100$>$ and $<$110$>$ show similar values as expected for a 12-spot pattern.
Below 120 K, the intensity for $<$100$>$ becomes larger than that for $<$110$>$ due to the change in the pattern from 12 spots to 4 spots. 
The intensities for both regions significantly decrease below 120 K, and become similar again below 60 K, which corresponds to the broadening of the 4 spots.
Therefore, while the triangular-square SkX transition occurs below 120 K in common with \eight, the square SkX is severely disordered below $\sim$ 90 K.
This is in accord with the disordering of the helical state of Co spins below $\sim$ 90 K on ZFC\cite{Karube_776} due to the development of short-range antiferromagnetic correlation of Mn spins. 

The radial $q$ dependence of the SANS intensity for $<$100$>$ is presented in Fig. 9(d). 
From 120 K to 60 K, where the triangular-square SkX transition occurs, the peak center exhibits a large shift from 0.06 nm$^{-1}$ to 0.08 nm$^{-1}$, and the magnitude and width of the peak significantly decreases and increases, respectively. 
Thus, the triangular-square structural transformation of SkX in \seven\ is also accompanied by the large increase in $q$ as well as in its width.
This indicates that the metastable SkX is heavily disordered, similarly to the helical state during ZFC [Fig. 3(l)].

\subsubsection{Field dependence of metastable SkX}

Next, we discuss how the metastable SkX state varies with field sweeping in the well-ordered region at 100 K (Fig. 10) and in the disordered region at 60 K (Fig. 11).

The field variation of the metastable SkX state at 100 K is presented in Fig. 10.
Figure 10(a) shows the field variation in the SANS patterns at 100 K after a field cooling (FC) at 0.025 T.
The detailed field dependence of the SANS intensity integrated over the region around $<$100$>$ and $<$110$>$ is displayed in Fig. 10(c).
With increasing the field to the positive direction, the initial 4-spot pattern changes to a uniform ring above 0.055 T, where the scattering intensities for $<$100$>$ and $<$110$>$ completely overlap while showing a clear shoulder, and the signal persists up to 0.1 T.
This change is ascribed to the transformation from the square SkX to an orientationally disordered triangular SkX similarly to the case of \eight\cite{Karube_884}.
In the field sweeping to the negative direction, the 4-spot pattern again changes to the ring-like one at $-$0.055 T but the intensity is much weaker than that observed at 0.055 T.
In the returning process from 0.15 T to 0 T, clear SANS signal is not observed because the metastable SkX is completely destroyed at a high-field region and a conical state is stabilized instead.
The large and asymmetric hysteresis of the intensity plot in Fig. 10(c) is attributed to the existence of the metastable SkX state.
For comparison, the field dependence of the SANS pattern and the integrated intensity at 100 K after zero-field cooling is shown in Fig. 10(b) and Fig. 10(d), respectively. 
In this case, 4 spots corresponding to the helical multi-domain state with $\vect{q}$ $\parallel$ $<$100$>$ is observed up to 0.05 T, but the signal disappears at 0.1 T without exhibiting a clear ring-like patten. 
This field variation for helical state after ZFC is totally different from that observed after the FC.
Therefore, it can be concluded that the metastable SkX created by the FC survives over a wide field region at 100 K, and the square SkX changes to the triangular SkX at high fields.

We also show the field variation of the metastable SkX state at 60 K in Fig. 11, and discuss another transition to the second equilibrium skyrmion phase.
The field-dependent SANS patterns at 60 K after the FC (0.025 T) are presented in Fig. 11(a).
The SANS intensity integrated over the region around $<$100$>$ and $<$110$>$ in this process is plotted against magnetic field in Fig. 11(c).
As the field is increased to the positive direction, the initial pattern with broad 4 spots originating from the disordered square SkX changes to a broad ring pattern above 0.07 T, resulting in the almost equal scattering intensities for $<$100$>$ and $<$110$>$.
In the field sweeping to the negative field region, the broad ring pattern with similar intensity is observed at $-$0.07 T and $-$0.1 T.
Therefore, the field variation after the FC is less asymmetric between the positive fields and the negative fields as compared with the result at 100 K.
In the returning process from the field-induced ferromagnetic phase ($\pm$0.2 T) down to 0 T, the broad ring pattern appears again and remains down to zero field. 
Note that the intensity of the ring pattern around zero field is relatively large and comparable to that of the initial broad 4-spot pattern just after the FC. 
For comparison, we show SANS patterns observed at the same magnetic fields after ZFC in Fig. 11(b), and the field dependence of the SANS intensity is plotted in Fig. 11(d).
In this process, the broad 4-spot pattern (disordered helical state) changes to a broad ring above 0.07 T similar to the FC case.
The field dependence commonly observed for the FC (both for positive and negative field directions) and ZFC processes is in accord with the existence of another field-induced equilibrium phase at low temperatures.
In our previous study on \seven, the broad ring pattern at low temperatures has been identified to be three-dimensionally disordered skyrmions, which are stabilized by frustrated magnetism of Mn spins and exist as an equilibrium phase that is distinct from the conventional SkX phase just below \Tc\cite{Karube_776}.
Therefore, the observed change from the broad 4-spot pattern to the broad ring pattern above 0.07 T in the FC case corresponds to the transition from the metastable square SkX state (originating from the high-temperature equilibrium SkX phase) to the other equilibrium disordered skyrmion phase.
The ring pattern remaining at zero field after the field decreasing process is explained in terms of the metastable skyrmions that are created initially in the high field region and persist down to zero field.

On the basis of the above results, the state diagram of the metastable SkX in \seven\ is summarized in the Fig. 2(e). The metastable square SkX (M-S-SkX) state exists below 120 K and at low fields inside the metastable triangular SkX (M-T-SkX) state. In addition, the equilibrium disordered skyrmion (E-DSk) phase is stabilized at low temperatures just above the reentrant spin glass transition around 30 K. 
Note that the E-DSk phase is discerned also in a negative-field region, reflecting its thermodynamical equilibrium nature.

\subsection{Summary of helical state and metastable skyrmion state}
Taking the SANS results for all the compounds into account, we come back to Fig. 2 and Fig. 3 again and discuss how the helical and the metastable skyrmion states change with Mn concentration.

\subsubsection{Correspondence between magnetization and $\vect{q}$ vector in helical state}
Table 1 summarizes the several quantities that characterize the helical state. 
One important effect of Mn substitution is the change in $\vect{q}$ vector direction: While the preferred orientations of $\vect{q}$ vectors are $<$111$>$ directions in \ten, they are $<$100$>$ directions in all the Mn-doped compounds. The helical structure is formed by the ferromagnetically coupled Co spins accompanied with DMI, but as described in Table 1 (see also Supplementary Fig. S1), saturation magnetization $M_\mathrm{s}$ measured at 2 K and 7 T takes a maximum at \nine. This suggests that Co and Mn are ferromagnetically coupled at least in the low Mn concentration region, as also demonstrated recently by X-ray magnetic circular dichroism (XMCD) measurements\cite{Ukleev_884}.

Mn substitution produces another important effect: Figures 3(e-h) summarize the temperature dependence of the helical $q$ value measured for all the compounds.
The $q$ value was determined as the peak center of Gaussian function fitted to the azimuthal-angle-averaged SANS intensity as a function of $q$.
In \ten, $q$ gradually and slightly increases by only $\sim$ 10\% upon cooling from \Tc\ to low temperatures.
On the other hand, \nine\ shows a large increase in $q$ below 50 K by $\sim$ 45\% while for \eight\ and \seven, $q$ also displays a large increase of $\sim$ 50\% below 120 K.
Notably, the large observed increases in $q$ almost coincide with the gradual decrease in $M$ from $T_\mathrm{H}$ to $T_\mathrm{L}$ as shown in Fig. 3(a-d).
The temperature region where $q$ significantly varies is indicated in the $T$-$x$ phase diagram in Fig. 1(b).

Figures 3(i-l) show the temperature dependence of the full width at half maximum (FWHM) of the Gaussian function fitted to the SANS intensity versus $q$, which provides a measure of the spatial coherence of the helimagnetic order.
The FWHM increases upon cooling in all the compositions while showing strong correlation with the increase in the magnitude of $q$ value [Figs. 3(e-h)]. 
Among them, \seven\ exhibits a significant increase below $\sim$ 90 K, which indicates that the helical state in \seven\ becomes severely disordered at low temperatures.
The increase in the FWHM for \nine\ and \eight\ also indicates the evolution of magnetic disorder to some extent, although they are less significant as compared with \seven.

The above temperature and Mn concentration dependence of helical states is well explained by the interplay between the ferromagnetically coupled Co spins forming the helical state and the antiferromagnetically coupled Mn spins.
Namely, the significant increase in the $q$ value (or decrease in the helical periodicity) at low temperatures is caused by the effective decrease in the ratio of the ferromagnetic exchange interaction to the DMI, which is attributed to short-range antiferromagnetic correlations of the Mn spins. These short-range correlations act as a source of disorder for helimagnetic Co spins, and start to develop at increasingly higher temperature (higher $T/T_\mathrm{c}$) as the Mn concentration is increased.
The increase of the $q$ value finally saturates below $T_\mathrm{L}$ probably because of the slowing of the antiferromagnetic fluctuations of Mn spins and resulting in a quasi-static disorder for the helimagnetic Co spins.
As the temperature is further reduced, Mn spins eventually freeze and undergo the reentrant spin glass transition as observed in the magnetization measurements for \eight\ and \seven\ [Figs. 3 (c) and (d)]. The freezing temperature $T_\mathrm{g}$ also increases with the Mn concentration due to the associated enhancement of the antiferromagnetic Mn spin correlations. 

\subsubsection{Summary of metastable skyrmion state}

Figure 2 is the summary of the equilibrium and metastable skyrmion phase diagrams on the $T$-$H$ plane determined by SANS and ac susceptibility measurements in \ten, \nine, \eight\ and \seven.
In all the compounds, the metastable SkX state is realized by a conventional field cooling via the equilibrium SkX phase just below \Tc\ and prevails over a very wide temperature and field region.
In addition, while the equilibrium and metastable skyrmions at high temperatures form a conventional triangular lattice described with triple-$\vect{q}$ vectors, the lattice structure transforms to the novel double-$\vect{q}$ states at low temperatures as follows. 
In \ten\ [Fig. 2(b)], the lattice form of the metastable SkX transforms from triangular to a rhombic one (M-R-SkX) with double-$\vect{q}$ $\parallel$ $<$111$>$ below $\sim$ 360 K with a broad coexistence region (a purple region). At all temperatures, the triangular lattice is restored as the field is increased.
In \nine\ [Fig. 2(c)], while the triangular lattice of the metastable skyrmions (M-T-SkX) persists over a wide temperature region, it transforms to square one (M-S-SkX) with double-$\vect{q}$ $\parallel$ $<$100$>$ below $\sim$ 50 K as shown with a pink region. The triangular lattice is also recovered at high fields. 
In \eight\ [Fig. 2(d)] and \seven\ [Fig. 2(e)], the triangular-square transition of skyrmion lattice occurs below $\sim$ 120 K, and the square SkX persists below the reentrant spin glass transition temperatures as plotted with yellow circles.
In \seven, the metastable square SkX state originating from the conventional SkX phase undergoes another transition to the other frustration-induced equilibrium skyrmion phase (E-DSk; orange region) as the field is increased at temperatures below $\sim$ 60 K.

The structural transitions of the skyrmion lattice to the rhombic one and to the square one are perhaps attributed to the enhanced magnetocrystalline anisotropy toward $\vect{q}$ $\parallel$ $<$111$>$ and $\vect{q}$ $\parallel$ $<$100$>$, respectively.
However, the $\vect{q}$ vector anisotropy alone is not sufficient to drive the lattice structural transition.
Importantly, these transformations are accompanied by an increase in absolute value of $q$: the triangular-rhombic transition in \ten\ is observed while $q$ value slightly increases over a broad temperature range [Fig. 3(e)], and the triangular-square transition in Mn-doped compounds occurs only when $q$ value significantly increases below a specific temperature of $T_\mathrm{H}$ [Fig. 3(f-h)]. 
As discussed quantitatively in the following, the increase in $q$ value is crucial for the transformation of the skyrmion lattice in terms of skyrmion density. 

The ratio of skyrmion density (number of the skyrmion per area) in the rhombic lattice ($n_\mathrm{R}$) to that in the triangular one ($n_\mathrm{T}$) is given by $n_\mathrm{R}/n_\mathrm{T} = \frac{3\sqrt{3}}{4\sqrt{2}}(a_\mathrm{T}/a_\mathrm{R})^2 = \frac{3\sqrt{3}}{4\sqrt{2}}(q_\mathrm{R}/q_\mathrm{T})^2$. Here, $a_\mathrm{T}$ ($q_\mathrm{T}$) and $a_\mathrm{R}$ ($q_\mathrm{R}$) are the lattice constant (the $q$ value) of the triangular lattice and the rhombic one, respectively.
Since the total number of metastable skyrmions that are topologically protected is conserved  during the transition from the triangular lattice to the rhombic one ($n_\mathrm{R}$ = $n_\mathrm{T}$), the lattice constant should decrease (the $q$ value should increase) by the factor of $\sqrt{\frac{4\sqrt{2}}{3\sqrt{3}}}$ $\sim$ 1.043 as the temperature is lowered. 
The $q_\mathrm{R}$ value that is consistent with this condition is presented with an dashed purple line in the $q(T)$ plot for \ten\ [inset of Fig. 3(e)]. 
Here, $q_\mathrm{T}$ is taken as the $q$ value at 412 K.
It is found that the observed small increase in $q$ value during the FC in \ten\ easily satisfies the above condition around $\sim$ 200 K.
This result is in accord with the fact that the transformation to rhombic lattice starts at high temperatures.

In the case of the transition to square lattice in the Mn-doped compounds, on the other hand, the aforementioned condition is more difficult to fulfill. 
The ratio of the skyrmion density in the square lattice ($n_\mathrm{S}$)  to that in the triangular one is given by $n_\mathrm{S}/n_\mathrm{T} = \frac{\sqrt{3}}{2}(a_\mathrm{T}/a_\mathrm{S})^2 = \frac{\sqrt{3}}{2}(q_\mathrm{S}/q_\mathrm{T})^2$, where $a_\mathrm{S}$ ($q_\mathrm{S}$) is the lattice constant (the $q$ value) of the square lattice.
To keep the skyrmion density constant ($n_\mathrm{S}$ = $n_\mathrm{T}$), the lattice constant should decrease (the $q$ value should increase) by the larger factor of $\sqrt{\frac{2}{\sqrt{3}}}$ $\sim$ 1.075. 
The value of $q_\mathrm{S}$ satisfying this condition is denoted with an dashed pink line in the $q(T)$ plot in \nine, \eight\ and \seven\ [Fig. 3(f-h)]. Here, $q_\mathrm{T}$ is taken as the $q$ value at the equilibrium SkX phase.
In \nine\ and \eight, $q$ value is almost $T$-constant at high temperatures. 
However, $q$ value significantly increases and exceeds the necessary $q_\mathrm{S}$ value below $T_\mathrm{H}$ ($\sim$ 50 K for \nine\ and $\sim$ 120 K for \eight\ and \seven), which would describe the transformation to the square lattice. 
This result well explains the observed onset temperature of the transformation to the square lattice.
Nevertheless, the observed increase in $q$ around $T_\mathrm{L}$ is as large as $q_\mathrm{S}/q_\mathrm{T}$ $\sim$ 1.5, and the skyrmion density becomes too large if we assume a uniformly and perfectly ordered square SkX as shown in Fig. 1(e).

Alternatively, the large increase in $q$ under the conserved skyrmion density can be reconciled with either of the following two scenarios: (i) a microscopic phase separation into a square SkX state and a helical phase as schematically illustrated in Fig. 1(f), or (ii) the onset of skyrmion deformation along the $<$100$>$ directions [Fig. 1(g)]. It should be noted that, in both (i) and (ii), the total number of skyrmions is identical to the original triangular SkX.
Although the latter deformed skyrmions have been observed in a thin-plate sample by LTEM\cite{Morikawa} and reproduced by a micromagnetic simulation\cite{Ukleev_884}, it is difficult to experimentally distinguish the two scenarios from SANS studies on a three-dimensional bulk sample. 
In any case, the large increase in $q$ value at low temperatures and the enhanced $\vect{q}$ vector anisotropy toward $\vect{q}$ $\parallel$ $<$100$>$ cooperatively drive the skyrmion lattice transformation as the metastable triangular SkX phase is cooled down to low temperatures. 

It is also noted that the square lattice of (elongated) skyrmions in the present case is distinct from a meron-antimeron square lattice with vanishing topological charge (genuine superposition of orthogonal double $\vect{q}$ vectors) as recently observed in a thin-plate sample of Co$_8$Zn$_9$Mn$_3$ by LTEM\cite{Yu_893}. 
The meron-antimeron state appears as an equilibrium state just before entering the equilibrium triangular SkX phase near \Tc.
While the meron-antimeron square lattice is stabilized by the in-plane shape anisotropy characteristic of a thin-plate specimen as theoretically predicted\cite{Yi,Lin}, the transition to the square SkX in the present case probably originates from magnetocrystalline anisotropy (local magnetization $\parallel$ $<$100$>$, which in turn results in $\vect{q}$ vector anisotropy along $<$100$>$) as well as the large increase in $q$, as discussed above.

\subsection{Lifetime of metastable skyrmion}

In the former sections, we revealed that the SkX state can persist at low temperatures by a field cooling process. 
Since the low temperature states are only metastable, the SkX should relax to the more stable conical state with some lifetime.
In our previous study (Ref. \cite{Karube_992}), we investigated temperature-dependent relaxation times of metastable skyrmions in \nine\ and observed extremely long lifetimes even at high temperatures. For a systematic understanding of behavior, in this section we extend the lifetime measurement to the other compounds \ten\ and \eight, and discuss how the metastable SkX lifetime varies with temperature and Mn concentration (Fig. 12 and Fig. 13). 

Figures 12(a-c) show $H$-$T$ phase diagrams of equilibrium states near \Tc\ in \ten, \nine\ and \eight, respectively.
For the purpose of better comparison between the three compounds, $T$ ranges are selected in the respective panels in such a way that normalized temperature $T/T_\mathrm{c}$ is from 0.90 to 1.01, as indicated on the upper abscissa. 
Clearly, the equilibrium SkX phase (green region) expands upon increasing the Mn concentration due to the increased chemical and magnetic disorder.
After a FC via the equilibrium SkX phase, the temporal variation of ac susceptibility $\chi^\prime (t)$ was measured at a fixed temperature, and this experiment was repeated at several different temperatures, as denoted with the colored circles.

The normalized ac susceptibility $\chi^\prime_\mathrm{N}(t) \equiv [\chi^\prime (\infty) - \chi^\prime (t)]/[\chi^\prime (\infty) - \chi^\prime (0)]$ is plotted as a function of time in Fig. 12(d-f).
Here, $\chi^\prime (0)$ and $\chi^\prime (\infty)$ correspond to an initial value for the metastable SkX state and the value for the equilibrium conical state as a fully relaxed state, respectively, and hence $\chi^\prime_\mathrm{N}$ = 1 for $t$ = 0 and $\chi^\prime_\mathrm{N}$ = 0 for $t$ $\rightarrow$ $\infty$.
In \ten\ [Fig. 12(d)], a clear relaxation from the metastable SkX state to the equilibrium conical state is observed at 404 K, just below the equilibrium SkX phase, and for which the relaxation time is the order of 10$^4$ s (several hours). 
The relaxation time further increases upon lowering the temperature. 
In \nine\ [Fig. 12(e)], the observed relaxation in the $\chi^\prime_\mathrm{N}(t)$ curve within the measurement time ($\sim$ 1 day) is less than 40\% even at 380 K, just below the equilibium SkX phase, and the relaxation becomes even longer as the temperature is lowered.
In \eight\ [Fig. 12(f)], only a few \% of relaxation is observed in the $\chi^\prime_\mathrm{N}(t)$ even at 280 K, just below the equilibrium SkX phase.
Therefore, the lifetime of metastable SkX at temperatures close to the equilibrium SkX phase boundary, for example the temperatures indicated with the red circles in Fig. 12(a-c), becomes relatively longer as the Mn concentration is increased.

To discuss more quantitatively, the $\chi^\prime_\mathrm{N}(t)$ curves are fitted to a stretched exponential function, $\exp \left\{ -(t/\tau)^\beta \right\}$.
The obtained $\beta$ values are in the range of 0.3 - 0.5, indicating a highly inhomogeneous distribution of relaxation times. 
The relaxation time $\tau$ in all the compounds (\ten, \nine\ and \eight) is plotted against $T/T_\mathrm{c}$ in Fig. 13(a).
The $\tau$ value exponentially increases as the temperature is lowered, and becomes virtually infinite when $T/T_\mathrm{c}$ is less than $\sim$ 0.9.
Remarkably, the data points from the three different compounds collapse onto a single curve although the applied magnetic field as well as the respective temperature regions are different. 
Following the arguments described in Ref. \cite{Oike}, all the data points are fitted to a modified Arrhenius law (pink solid line), $\tau = \tau_0 \exp \left\{ a(T_\mathrm{c} - T)/T\right\}$. 
Here, the activation energy for the relaxation from the metastable SkX state to the equilibrium conical state, as schematically illustrated in Fig. 13(b), is assumed to be $T$-dependent as $E_\mathrm{g} = a(T_\mathrm{c}-T)$ near \Tc, instead of constant $E_\mathrm{g}$ for standard Arrhenius law. 
The obtained fitting parameters are $a$ = 215 and $\tau_0$ = 63 s.
For comparison, reported values of $a$ and $\tau_0$ for several materials\cite{Oike,Birch} are summarized in Table 2.
Recently, Wild \textit{et al}. reported that $\tau_0$ sensitively depends on the applied magnetic field in their LTEM studies of a thin plate of Fe$_{1-x}$Co$_x$Si\cite{Wild}. 
For the present ac susceptibility measurements on bulk crystals of Co-Zn-Mn alloys, we used the applied magnetic field for each compound where the SANS intensity from the equilibrium SkX is strongest in the field sweeping measurement. The fact that the $\tau$ values from three different compounds collapse onto a single curve (namely, $\tau_0$ are the same) may originate from the fact that the applied field values are close to the optimal ones in the respective samples.

From the obtained value of $a$, the activation energy $E_\mathrm{g}$ is estimated to be larger than 8000 K at a temperature $T$ = 0.9\Tc\ for \ten\ and \nine.
This energy scale protecting the metastable skyrmion state is much larger than the ferromagnetic exchange interaction (several 100 K), and thus attributed to the topological nature of the skyrmion with a large diameter ($\sim$ 100 nm) that involves a great number of spins.

The coefficient of relaxation time $\tau_0$ is inversely correlated to the critical cooling rate\cite{Kagawa_AdvMat} for quenching the SkX phase to lower temperatures. Since the obtained value of $\tau_0$ in Co-Zn-Mn alloys is the order of a minute, the conventionally slow cooling rate ($dT/dt$ $\sim $ $-$1 K/min) is sufficient to quench the SkX phase.
This is quite different from the value of $\tau_0$ $\sim$ 10$^{-4}$ s in MnSi\cite{Oike}, where an ultra-rapid cooling rate ($dT/dt$ $\sim $ $-$100 K/s) is necessary to quench the SkX phase.
The large difference in $\tau_0$ that depends on the material is probably attributed to randomness in the system.
In the case of Co-Zn-Mn alloys, there are random site occupancies in the crystal structure; the 8$c$ site is randomly occupied by Co and Mn, and the 12$d$ site is occupied by Co, Zn, and Mn\cite{Hori,Xie,Bocarsly,Nakajima_CoZnMn}. 
This gives rise to ``weak pinning" in the terminology of density wave physics\cite{Gruner}, which may play an important role in the robust metastability of the skyrmion.

From a microscopic viewpoint, the destruction of metastable skyrmions in the bulk takes place through the creation of a pair of Bloch points, or equivalently an emergent magnetic monopole-antimonopole pair, from a singularity point of a skyrmion string, followed by their propagation\cite{Milde, Kagawa}, as schematically illustrated in Fig. 13(c).
Coefficients $a$ and $\tau_0$ are roughly governed by the creation and the propagation processes of monopole-antimonopole pairs, respectively.
In a clean system like MnSi, the monopole and antimonopole can easily move, which leads to skyrmion string destruction after the pair creation. 
In a dirty system like Co-Zn-Mn alloys, the movement of monopole and antimonopole is hindered by magnetic impurities or defects, and consequently $\tau_0$ of the metastable skyrmion string is significantly increased.
A similar long-lived metastable SkX that is accessible by a moderate cooling rate has been reported in Fe$_{1-x}$Co$_x$Si alloys\cite{Munzer,Milde, Wild}, which may bear some resemblance to the present case.
More recently, the increased lifetime of metastable SkX has been observed also in Zn-doped Cu$_2$OSeO$_3$, where $\tau_0$ in a Zn 2.5\% doped sample is 50 times larger than that in a non-doped sample while $a$ value is unchanged\cite{Birch}. 
In the present case of Co-Zn-Mn alloys, the good scaling of the $\tau$ vs $T/T_\mathrm{c}$ plot in Fig. 13(a) indicates that both $a$ and $\tau_0$ are almost independent of the Mn concentration.
This is probably because \ten\ already possesses substantial randomness in the site occupancy at 12$d$ site (2 Co and 10 Zn per unit cell), and thus the attempt time $\tau_0$ is already sufficiently long and not further increased by additional randomness due to the Mn substitution.
Nevertheless, Mn substitution expands the equilibrium SkX phase toward lower temperature as seen in Fig. 12(a-c) and Fig. 13(a), and hence the relaxation time just below the equilibrium SkX phase becomes longer. 
It is also interesting to note that the equilibrium skyrmion phase is expanded by the static magnetic disorder while thermal fluctuation is considered to be important for the realization of the equilibrium phase close to \Tc.

\section{Conclusion}

In the present study, to provide the perspective on the chiral magnetism in $\beta$-Mn-type Co-Zn-Mn alloys with bulk Dzyaloshinskii-Moriya interaction (DMI), we have performed magnetization, ac susceptibility and small-angle neutron scattering measurements on single-crystal samples of \coznmn\ with $x$ = 0, 2, 4 and 6.
The Mn-free end member \ten\ exhibits a helimagnetic ground state (periodicity $\lambda$ $\sim$ 156 nm) below the transition temperature \Tc\ $\sim$ 414 K, where the helical propagation vector $\vect{q}$ is aligned to $<$111$>$ at low temperatures (Fig. 5).  
Upon applying magnetic fields, \ten\ exhibits an equilibrium SkX phase above 400 K in a narrow temperature and magnetic field region (Fig. 4), which is quenched down to lower temperatures as a metastable state by a conventionally-slow field cooling (Fig. 6). 
The lifetime of the metastable SkX is extremely long and virtually infinite below 380 K (Figs. 12 and 13).
The metastable SkX state is highly robust and persists over the whole temperature range below \Tc\ and a wide magnetic field region, including room temperature and zero field (Fig. 2).
The lattice of metastable skyrmions distorts and transforms from a conventional triangular one to a rhombic one at low temperatures and low magnetic fields (Figs. 2 and 6).

As the partial substitution with Mn proceeds, \Tc\ decreases and the preferred $\vect{q}$ orientation switches to $<$100$>$. 
The saturation magnetization is the largest for \nine\ (Table 1).
At low temperatures, the helical $q$ value significantly increases, or equivalently $\lambda$ significantly decreases, below $T_\mathrm{H}$ $\sim$ 50 K for \nine\ and below $T_\mathrm{H}$ $\sim$ 120 K for \eight\ and \seven\ (Fig. 3). 
In \seven, the helical state is severely disordered at low temperatures.
Upon further decreasing temperature, a reentrant spin glass transition occurs at $T_\mathrm{g}$ $\sim$ 10 K for \eight\ and $T_\mathrm{g}$ $\sim$ 30 K for \seven\ while such a transition is not observed for \ten\ and \nine\ down to 2 K (Figs. 1 and 3). 
In common with \ten, a long-lived metastable SkX state is realized in the Mn-doped materials by a moderate FC through the equilibrium SkX phase, and persists over a wide temperature and field region. 
On the other hand, the lattice form of the metastable SkX changes to a square one at low temperatures (Figs. 2, 7, 8 and 9). 
While the triangular SkX is dictated by the applied field as $\vect{q}$ $\perp$ $H$, the square SkX is governed by the magnetocrystalline anisotropy as $\vect{q}$ $\parallel$ $<$100$>$ regardless of the applied field direction (Figs. 7 and 8). 
During the transition to the square lattice, the periodicity of SkX significantly shrinks below $T_\mathrm{H}$ similar to the helical periodicity in zero-field cooling (Fig. 3).

From these results, we conclude the followings: The helical and skyrmion states in Co-Zn-Mn alloys are basically formed by ferromagnetic Co spins in the presence of the DMI. 
While Co and Mn are ferromagnetically coupled at least in the low Mn concentrations, antiferromagnetic Mn-Mn correlations become increasingly significant in the higher Mn concentrations and start to develop at higher temperatures. As the temperature is lowered, the development of antiferromagnetic Mn-Mn correlation leads to a disordering of the helical state and simultaneously decreases helical pitch, and ultimately undergoing a spin freezing transition at very low temperatures. 
The robust metastability of skyrmions is attributed to the topological protection by a large number of involved Co spins as well as weak pinning from the magnetic disorder.
The structural transformations between metastable triangular SkX and either a rhombic one or a square one are driven by a decrease in the distance between the skyrmions under the influence of the magnetocrystalline anisotropy that favors $\vect{q}$ $\parallel$ $<$111$>$ and $\vect{q}$ $\parallel$ $<$100$>$ in undoped and Mn-doped compounds, respectively.
These findings unveil the complex interplay between chiral magnetism and the frustrated Mn spins that is also greatly affected by magnetic disorder and anisotropy, and provide a significant understanding of the topological phases and properties in this class of $\beta$-Mn-type chiral magnets.

\begin{acknowledgments}
We are grateful to T. Arima, T. Nakajima, D. Morikawa, X. Z. Yu, L. C. Peng and N. Nagaosa for fruitful discussions. 
We thank M. Bartkowiak for support of SANS experiments above room temperature at Paul Scherrer Institute (PSI), Switzerland.
This work was supported by JSPS Grant-in-Aids for Scientific Research (Grant No. 24224009 and No. 17K18355), JST CREST (Grant No. JPMJCR1874), the Swiss National Science Foundation (SNSF) Sinergia network `NanoSkyrmionics (Grant No. CRSII5\_171003)', the SNSF projects 200021\_153451, 200021\_188707 and 166298, and the European Research Council project CONQUEST.

The datasets for the SANS experiments done on D33 at the ILL are available through the ILL data portal (http://doi.ill.fr/10.5291/ILL-DATA.5-42-410 and http://doi.ill.fr/10.5291/ILL-DATA.5-42-443). 
\end{acknowledgments}

\bibliographystyle{apsrev}

\begin{thebibliography}{99}
\bibitem{S-SkX} As explained in Section III-F-2, there are two possibilities for the observed double-$\vect{q}$ state with increased $q$ values: One is a coexistence of the square lattice of skyrmions and the helical state [Fig. 1(f)], and the other is elongated-skyrmion state on the square lattice [Fig. 1(g)]. Hereafter, we refer to the double-$\vect{q}$ state as ``square lattice of skyrmions (or square SkX)" for the purpose of simplicity throughout the paper.
\bibitem{Note_RSG} It should be noted that the helimagnetic order formed by Co spins and the spin glass state of frustrated Mn spins coexist. Therefore, it may also be called ``embedded spin glass" as used in Ref. \cite{Ronnow}.
\bibitem{Footnote_776} In Ref. \cite{Karube_776} we reported for \seven\ that the real part of the ac susceptibility exhibits a drop, the imaginary part a peak at $T_\mathrm{g}$ $\sim$ 30 K, and the frequency-dependent $T_\mathrm{g}$ increases (29 K $\rightarrow$ 34 K) as the ac frequency $f$ is increased (0.1 Hz $\rightarrow$ 1000 Hz). The frequency-dependence of $T_\mathrm{g}$ satisfies a power law $f^{-1} = \tau_0[T_\mathrm{g}(f)/T_\mathrm{g}(0)-1]^{-z\nu}$ ($\tau_0$ = 5.3 $\times$ 10$^{-8}$ s, $T_\mathrm{g}(0)$ = 24.6 K, $z\nu$ = 10.7). The slow spin flipping time $\tau_0$ $\sim$ 10$^{-8}$ s is a typical value for reentrant spin glass\cite{Hanasaki}. A similar measurement for \seven\ was also reported in Ref. \cite{Bocarsly}.
\bibitem{Footnote_1010} The asymmetric peak broadenings around the [-111] and [1-1-1] directions at 1.5 K are attributed to imperfect summations of SANS patterns along the rocking direction.
\bibitem{Demag} On the basis of demagnetization coefficients in the rectangular-shaped sample and a magnetization value at 295 K, the extra tilt angle $\alpha$ of the effective magnetic field $H_\mathrm{eff}$ from the external field $H$ was calculated to be $\alpha$ = 23$^\circ$, showing a good agreement with the observed peak position of the rocking curve at 295 K.
\end{thebibliography}

\newpage

%%%%%%%%%%%%%%%%%  FIG1  %%%%%%%%%%%%%%
\begin{figure}[htbp]
\begin{center}
\includegraphics[width=15.5cm]{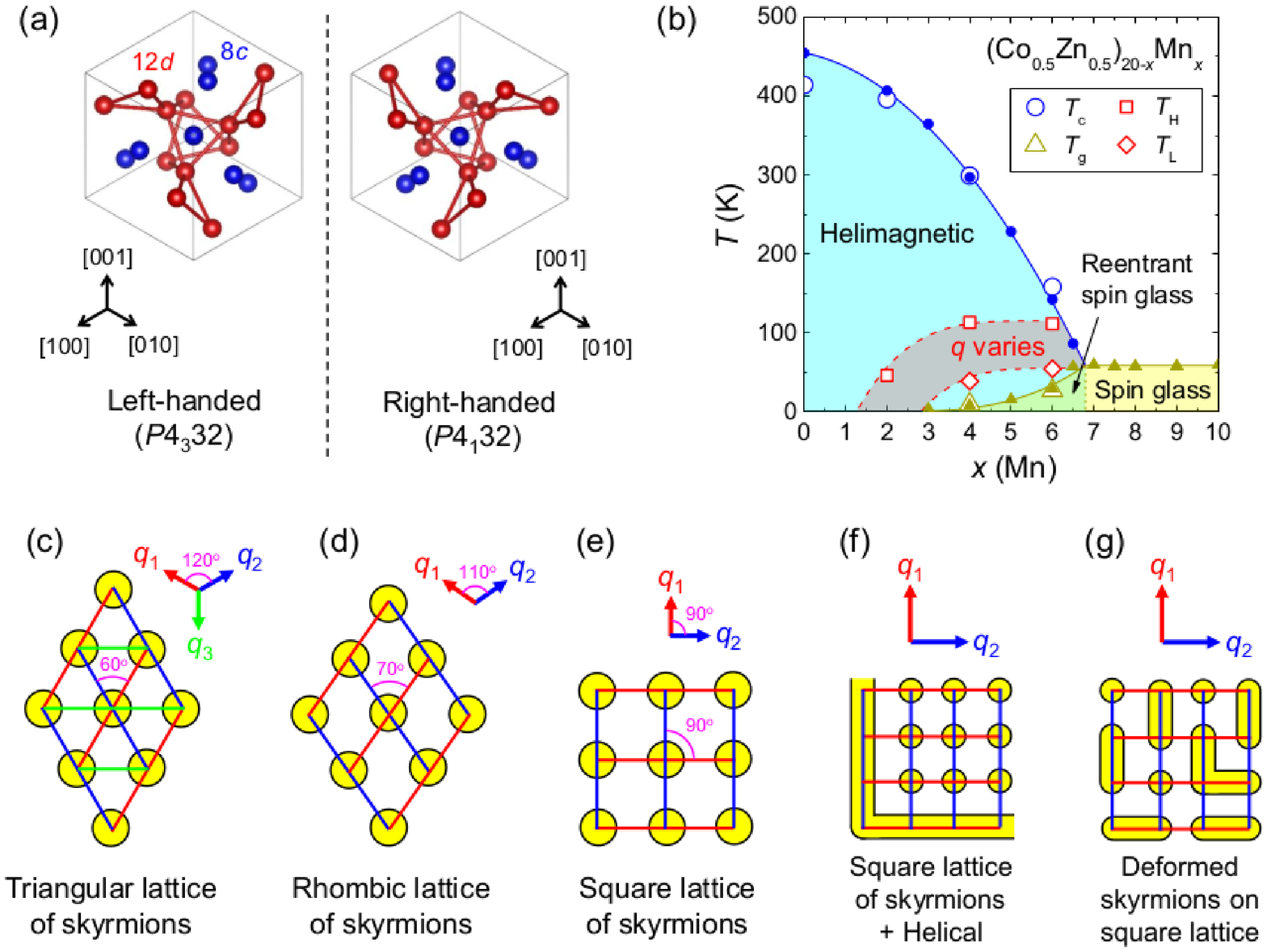}
\end{center}
\end{figure}
%%%%%%%%%%%%%%%%%  FIG1  %%%%%%%%%%%%%%
\noindent
FIG. 1. (a) Schematics of $\beta$-Mn-type chiral crystal structures as viewed along the [111] direction. Two enantiomers with the space group $P$4$_3$32 (left-handed structure) and $P$4$_1$32 (right-handed structure) are shown. Blue and red circles represent 8$c$ and 12$d$ Wyckoff sites, respectively. The network of 12$d$ sites forms a hyper-kagome structure composed of corner-sharing triangles. (b) Temperature ($T$) - Mn concentration ($x$) phase diagram in (Co$_{0.5}$Zn$_{0.5}$)$_{20-x}$Mn$_x$ (0 $\leq$ $x$ $\leq$ 10) at zero field, determined by magnetization ($M$) measurements. Closed symbols are data of polycrystalline samples reproduced from our previous work in Ref. \cite{Karube_776}. Copyright 2018, American Association for the Advancement of Science. Open symbols are data of single-crystalline samples ($x$ = 0, 2, 4, 6) in the present study [see Fig. 3(a-d) for the detailed determination of the phase boundaries]. 
A shaded region with red indicates the temperature range $T_\mathrm{L}$ $\leq$ $T$ $\leq$ $T_\mathrm{H}$ where $M$ at 20 Oe and the magnitude of helical wavevector ($q$) vary (decreases and increases respectively on cooling) significantly. 
(c-g) Schematic illustrations of various skyrmion lattices: (c) triangular lattice, (d) rhombic lattice, (e) square lattice, (f) coexistence of square lattice of skyrmions and helical state, (g) I- or L-like deformed skyrmions on a square lattice. Corresponding hybridized $\vect{q}$ vectors are also presented in each panel. 
\\

%%%%%%%%%%%%%%%%%  FIG2  %%%%%%%%%%%%%%
\begin{figure}[htbp]
\begin{center}
\includegraphics[width=10cm]{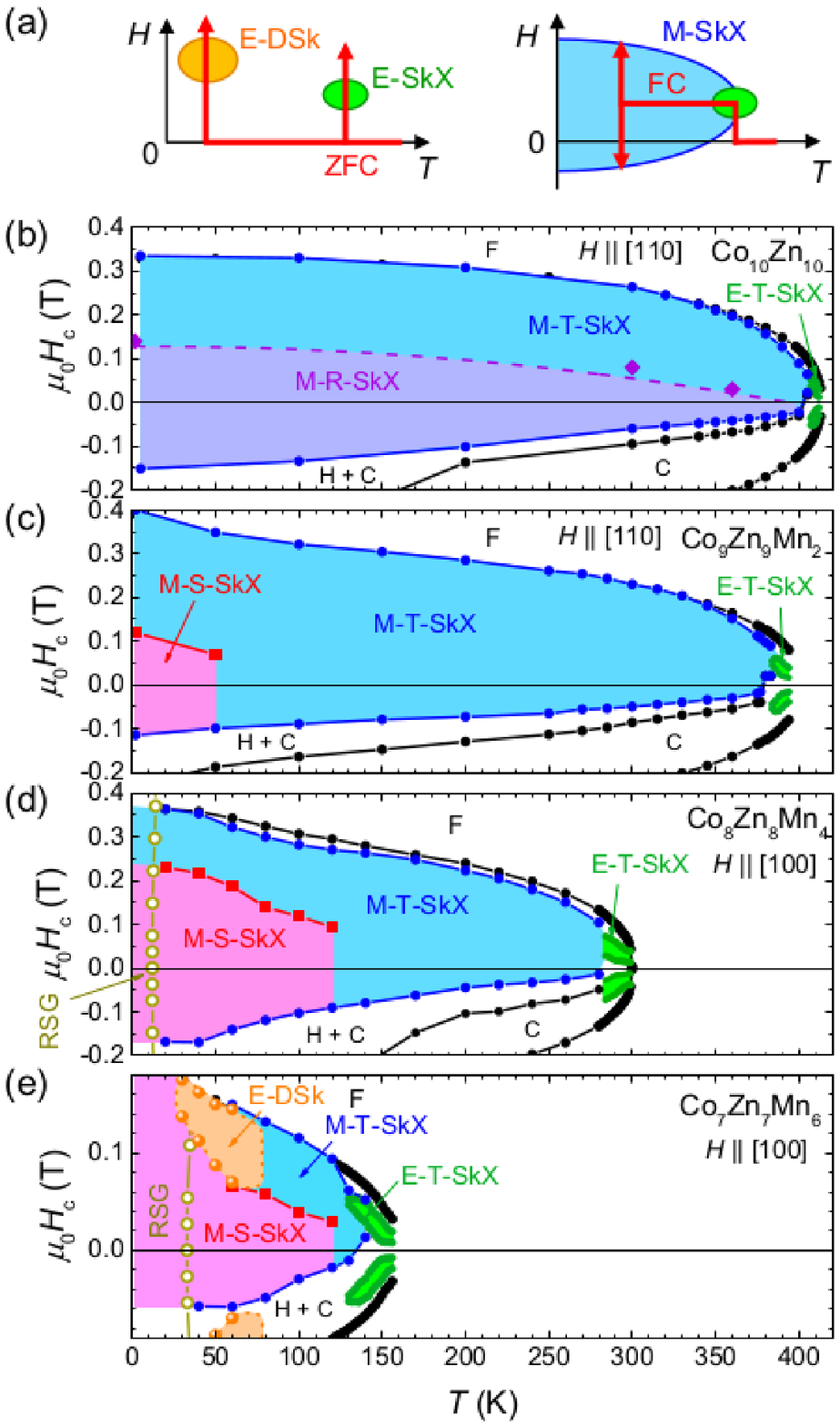}
\end{center}
\end{figure}
%%%%%%%%%%%%%%%%%  FIG2  %%%%%%%%%%%%%%
\noindent
FIG. 2. Summary of equilibrium and metastable skyrmion states in temperature ($T$) - magnetic field ($H$) plane in (b) \ten, (c) \nine, (d) \eight\ (reproduced from our previous work in Ref. \cite{Karube_884}. Copyright 2016, Springer Nature) and (e) \seven.
The measurement processes for the state diagrams are schematically illustrated in panel (a). The equilibrium skyrmion crystal (SkX) phase just below \Tc\ (green area) and the equilibrium disordered skyrmion (DSk) phase (orange area) are determined by field-sweeping measurements of ac susceptibility and SANS after zero-field cooling (ZFC). The metastable SkX state [triangular lattice (light-blue area), rhombic lattice (purple area) and square lattice (pink area)] are determined by field-sweeping measurements toward positive and negative directions after the field cooling (FC) via the positive-field equilibrium SkX phase. 
Here, we use the following notations; H: helical, C: conical, F: ferromagnetic, E: equilibrium, M: metastable, T: triangular, R: rhombic, S: square, SkX: skyrmion crystal, DSk: disordered skyrmions, and RSG: reentrant spin glass. 
The phase boundaries are determined by ac susceptibility ($\chi^\prime$). 
For the purpose of comparison with SANS measurements, calibrated values ($H_\mathrm{c}$) are used for the magnetic field. 
The details about the determination of the phase boundaries of the equilibrium SkX state and the metastable SkX state in \ten\ are described in the captions of Fig. 4(d) and Supplementary Fig. S2(c), respectively. 
The boundaries between the metastable triangular SkX and the metastable rhombic SkX (purple diamonds) are determined by SANS results [see Fig. 6(d) and Supplementary Figs. S2(d) and S3(d) for the details].
The boundary between the metastable triangular SkX and the metastable square SkX (red squares) are determined as inflection points of $\chi^\prime(H)$ as detailed in Ref. \cite{Karube_884}.
The reentrant spin glass transition temperatures (yellow open circles) are determined as inflection points of a sharp drop in $\chi^\prime(T)$, as also observed in $M(T)$ in the zero-field-cooled field-warming process [Fig. 3(c, d)].
The equilibrium DSk phase in \seven, determined as the region where a spherical SANS pattern is observed, was reproduced from our previous work in Ref. \cite{Karube_776}. Copyright 2018, American Association for the Advancement of Science.
\\

\newpage

%%%%%%%%%%%%%%%%%  FIG3  %%%%%%%%%%%%%%
\begin{figure}[htbp]
\begin{center}
\includegraphics[width=16cm]{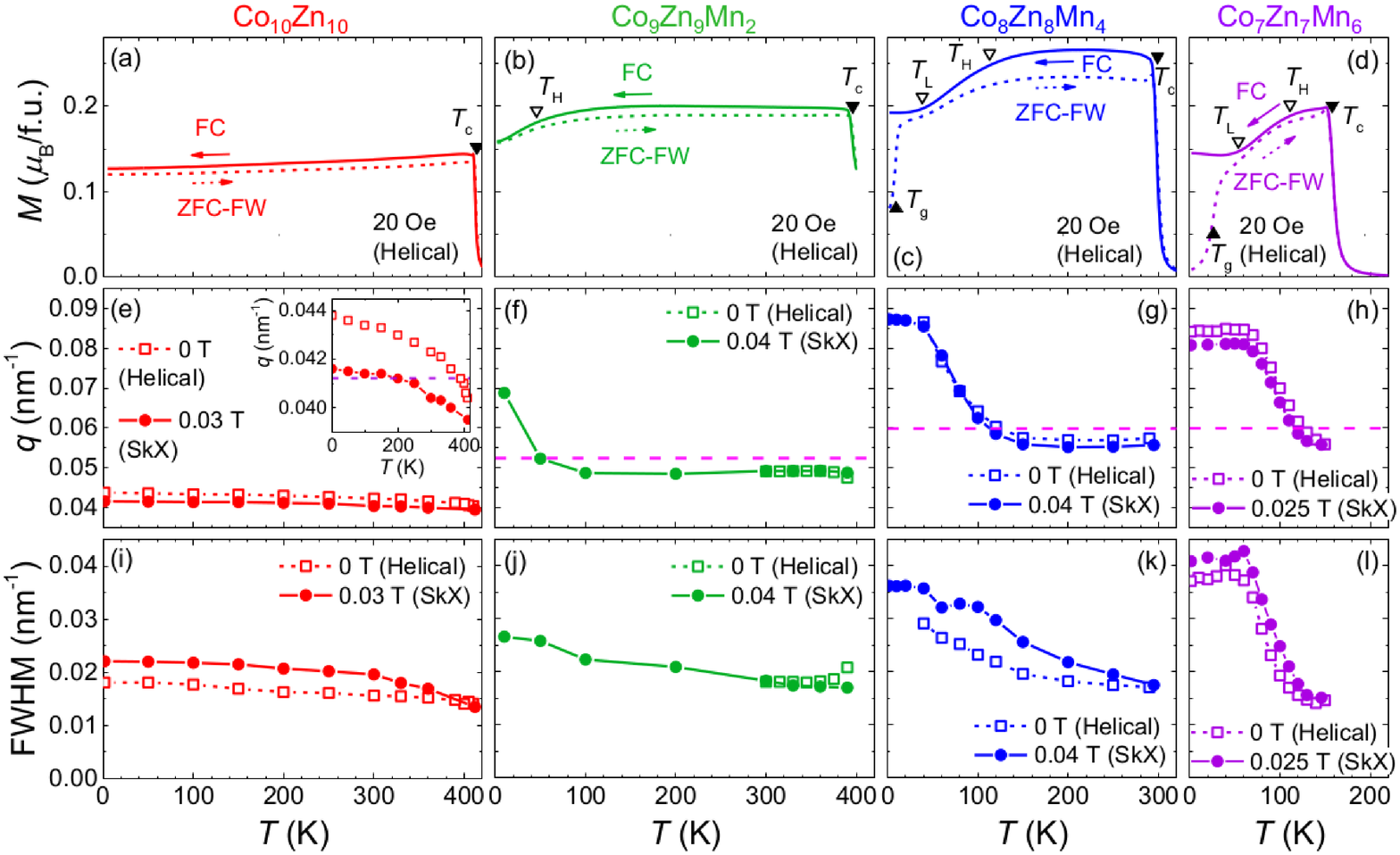}
\end{center}
\end{figure}
%%%%%%%%%%%%%%%%%  FIG3  %%%%%%%%%%%%%%
\noindent
FIG. 3. 
Temperature dependence of magnetization and $\vect{q}$ vector in \ten\ (red), \nine\ (green), \eight\ (blue) and \seven\ (purple).
(a-d) Magnetization ($M$) under a magnetic field of 20 Oe is plotted against temperature. 
Solid lines show the data collected during field cooling (FC), and broken lines represent those taken in a field-warming run after a zero-field cooling down to 2 K (ZFC-FW). 
The helimagnetic transition temperature (\Tc) indicated with closed triangles is determined as an inflection point of a sharp increase in $M$ on the 
FC process.
$T_\mathrm{H}$ and $T_\mathrm{L}$ are the temperatures at which a gradual decrease in $M$ on cooling starts and ends, respectively. 
These temperatures are determined as inflection points with a broad peak in $dM/dT$ on the FC process and presented with open triangles.
The reentrant spin glass transition temperature ($T_\mathrm{g}$) denoted with closed triangles is determined as an inflection point of a sharp increase in $M$ on the ZFC-FW process. 
These temperatures are also plotted in the $T$-$x$ phase diagram in Fig. 1(b).
(e-h) Temperature variation of the magnitude of $q$ in the helical state at zero field (open squares), and $q$ in equilibrium and metastable SkX states under magnetic fields (closed circles). 
The $q$ values are determined as the peak center of Gaussian function fitted to SANS intensity as a function of $q$. 
The inset of panel (e) shows an enlarged view of $q(T)$ for \ten.
The purple dashed line in the inset of (e) and the pink dashed line in (f-h) indicate the expected $q$ values for rhombic and square lattice of skyrmions (= 1.043$q_\mathrm{T}$ and 1.075$q_\mathrm{T}$, where $q_\mathrm{T}$ are the values in the equilibrium triangular SkX state), respectively, with the assumption of constant skyrmion density from the high-temperature triangular lattice (see discussion in Section III-F-2 for the details).
(i-l) Temperature dependence of full width at half maximum (FWHM) of the Gaussian function fitted to SANS intensity versus $q$
\\

%\newpage

%%%%%%%%%%%%%%%%%% TABLE1  %%%%%%%%%%%%%%
\begin{table}[htbp]
\begin{center}
\begin{tabular}{c|ccccc} \hline 
\:\:\: Composition \:\:\: & \:\:\: \ten \:\:\: & \:\:\: \nine \:\:\: & \:\:\: \eight \:\:\: & \:\:\: \seven \:\:\: \\ \hline 
\:\:\: $T_\mathrm{c}$ (K) \:\:\: & 414 & 396 & 299 & 158 \\ 
\:\:\: $M_\mathrm{s}$ ($\mu_\mathrm{B}$/f.u.) \:\:\: & 11.3 & 14.3 & 12.7 & 7.97 \\ 
\:\:\: $\lambda_\mathrm{max}$ (nm) \:\:\: & 156 & 132 & 110 & 112 \\ 
\:\:\: $\lambda_\mathrm{min}$ (nm) \:\:\: & 143 & 91 & 73 & 74 \\ 
\:\:\: Preferred $\vect{q}$ direction \:\:\: & $<$111$>$ & $<$100$>$ & $<$100$>$ & $<$100$>$ \\ \hline
\end{tabular}
\end{center}
\end{table}
%%%%%%%%%%%%%%%%%%  TABLE1  %%%%%%%%%%%%%%
\noindent
TABLE. 1. Summary of several physical parameters for helimagnetic state in \ten, \nine, \eight\ and \seven. 
Helimagnetic transition temperature $T_\mathrm{c}$ is determined from the temperature variation in magnetization under 20 Oe. 
Saturation magnetization $M_\mathrm{s}$ is defined as a magnetization value at 2 K and 7 T. Helimagnetic periodicity $\lambda$ (maximum value $\lambda_\mathrm{max}$ at high temperature and minimum value $\lambda_\mathrm{min}$ at low temperature) is calculated from the $q$ value obtained from SANS measurements in zero field. 
The preferred orientation of the helical $\vect{q}$ vector is also determined from SANS measurements. 
\\

\newpage

%%%%%%%%%%%%%%%%%  FIG4  %%%%%%%%%%%%%%
\begin{figure}[htbp]
\begin{center}
\includegraphics[width=16cm]{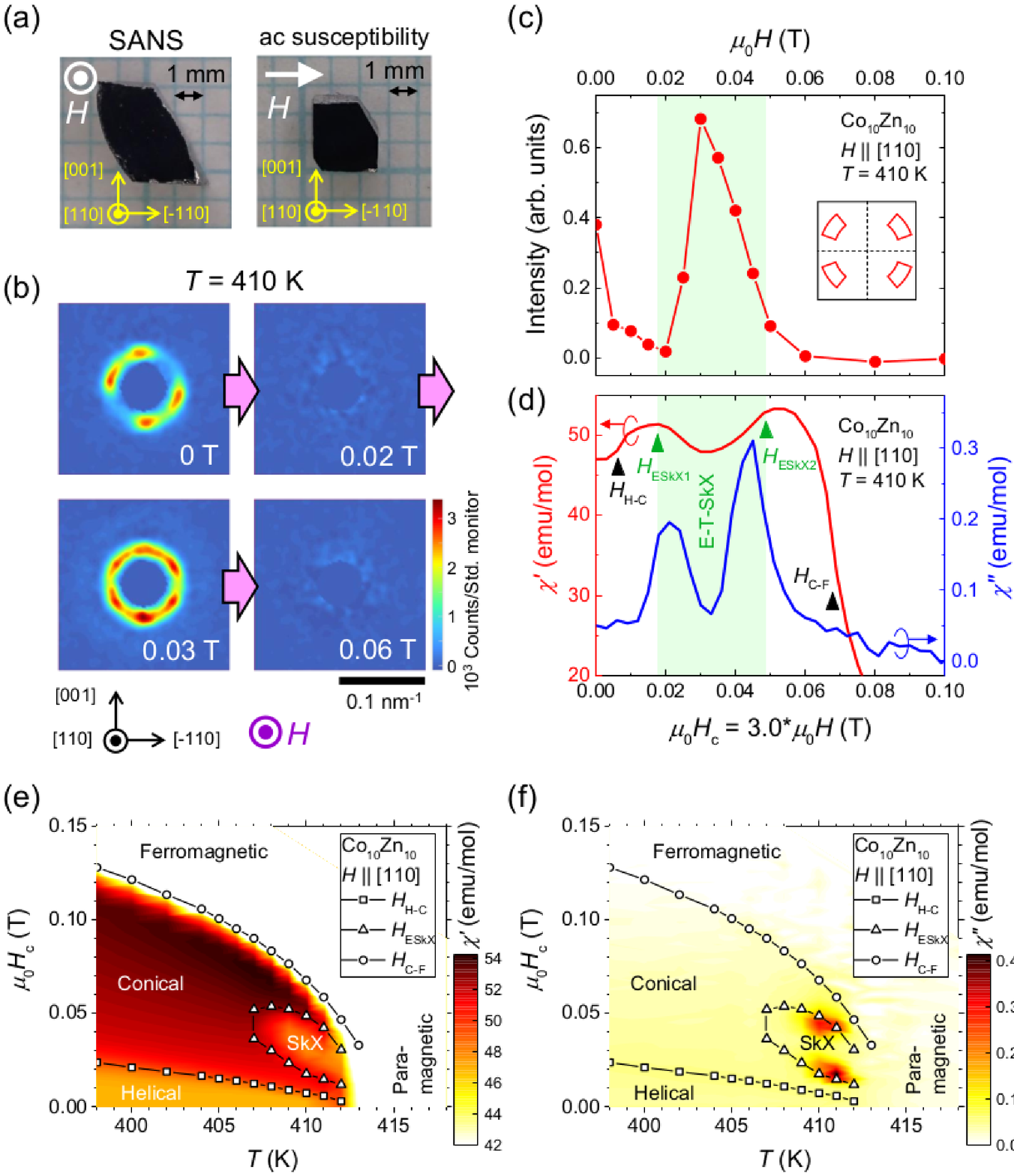}
\end{center}
\end{figure}
%%%%%%%%%%%%%%%%%  FIG4  %%%%%%%%%%%%%%
\noindent
FIG. 4. Identification of equilibrium SkX phase in \ten. (a) Photos of single-crystalline bulk samples of \ten\ used for SANS and ac susceptibility measurements. Crystal axes and the direction of applied magnetic field ($H$) are also indicated. 
(b) SANS images at selected fields in the field increasing run at 410 K. The intensity scale of the color plot is fixed for the 4 panels. 
(c) Field dependence of SANS intensity at 410 K. The intensity was integrated over the region close to $<$111$>$ (red area in the inset) as detailed in Fig. 5(c).
(d) Field dependence of the real part ($\chi^\prime$, red line) and the imaginary part ($\chi^{\prime\prime}$, blue line) of the ac susceptibility in the field increasing run at 410 K. Since the demagnetization factor is different from that in the SANS measurement, the field values are calibrated as $H_\mathrm{c} = 3.0 \times H$. 
The phase boundary between the helical and conical states ($H_\mathrm{H-C}$, black triangle) is determined as the inflection point of $\chi^\prime$. The phase boundaries between the equilibrium SkX and conical states ($H_\mathrm{ESkX1(2)}$, green triangles) are determined as the peak positions at both sides of the dip structure in $\chi^\prime$. The phase boundary between the conical and induced-ferromagnetic states ($H_\mathrm{C-F}$, black triangle) is determined as the inflection point of $\chi^\prime$. 
The equilibrium SkX region is indicated with the light-green shading in panels (c) and (d). 
(e) Contour plot of $\chi^\prime$ on the $T$ - $H_\mathrm{c}$ plane. 
The region with $\chi^\prime$ $\leq$ 42 emu/mol is displayed with white color.
(f) Contour plot of $\chi^{\prime\prime}$ on the $T$ - $H_\mathrm{c}$ plane. Phase boundaries determined by $\chi^\prime$ [panel (d)] and similar data points at different temperatures are plotted as open symbols in panels (e) and (f). 
\\

\newpage

%%%%%%%%%%%%%%%%%  FIG5  %%%%%%%%%%%%%%
\begin{figure}[htbp]
\begin{center}
\includegraphics[width=14cm]{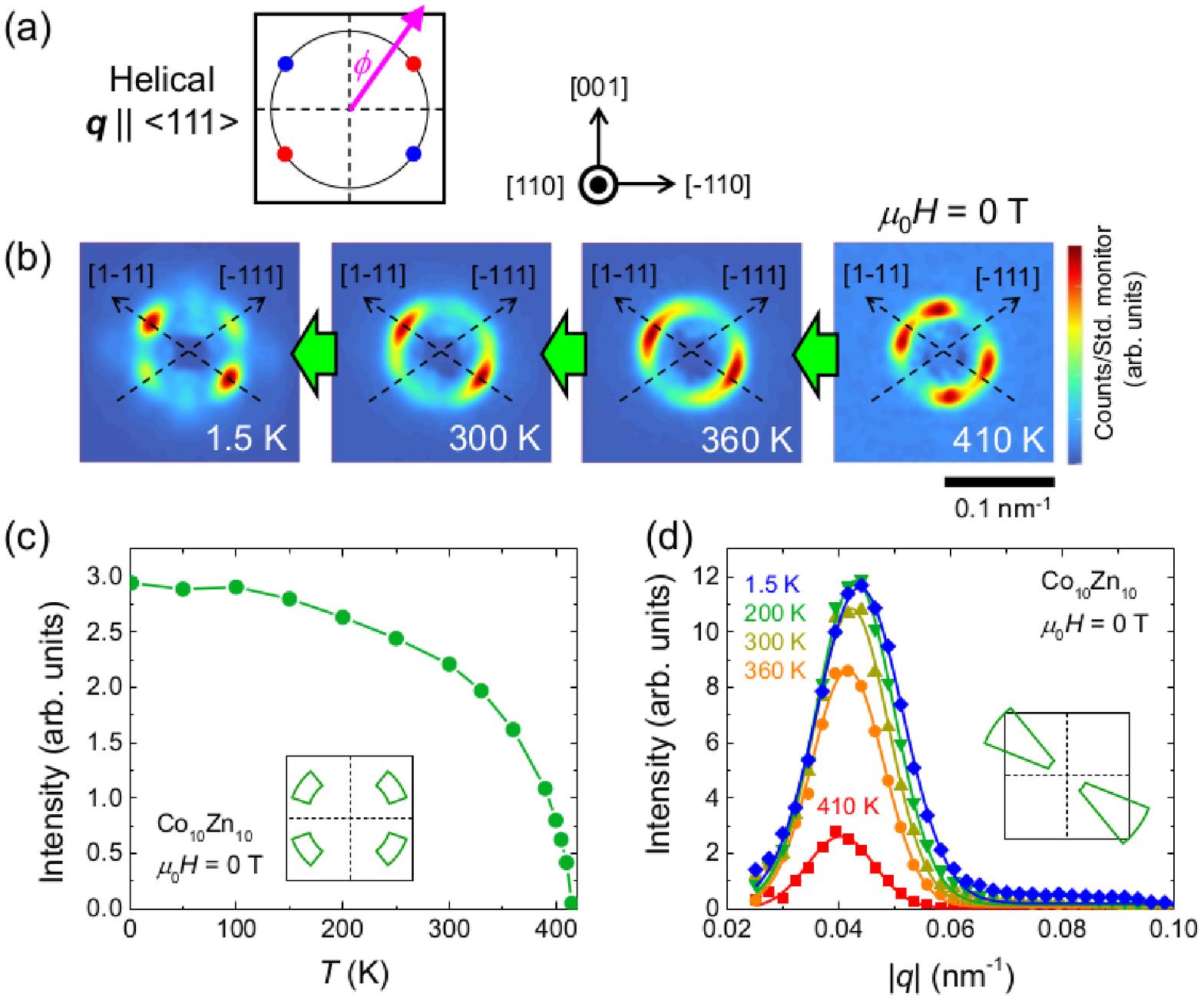}
\end{center}
\end{figure}
%%%%%%%%%%%%%%%%%  FIG5  %%%%%%%%%%%%%%
\noindent
FIG. 5. Temperature dependence of helical state in \ten\ in zero field. (a) Schematic SANS pattern on the (110) plane expected for a helical state with $\vect{q}$ $\parallel$ $<$111$>$. The helical state forms four domains with $\vect{q}$ $\parallel$ [-111] (2 red spots), [1-11] (2 blue spots), [11-1] (out of the plane) and [111] (out of the plane), respectively, resulting in 4 spots on the (110) plane at $\phi$ = 55$^\circ$, 125$^\circ$, 235$^\circ$, 305$^\circ$. Here, $\phi$ is defined as the clockwise azimuthal angle from the vertical direction. (b) SANS images observed at 410 K, 360 K, 300 K and 1.5 K. The intensity scale of the color plot varies between each panel. The [-111] and [1-11] directions are indicated with broken arrows. (c) Temperature dependence of the SANS intensity integrated over the azimuthal angle area at $\phi$ = 55$^\circ$, 125$^\circ$, 235$^\circ$, 305$^\circ$ with the width of $\Delta\phi$ = 30$^\circ$  (green area in the inset). (d) Radial $|q|$ dependence of SANS intensity, integrated over the azimuthal angle area at $\phi$ = 125$^\circ$, 305$^\circ$ with the width of $\Delta\phi$ = 30$^\circ$ (green area in the inset), at several temperatures. The data points are fitted to a Gaussian function (solid line). 
\\

\newpage

%%%%%%%%%%%%%%%%%  FIG6  %%%%%%%%%%%%%%
\begin{figure}[htbp]
\begin{center}
\includegraphics[width=14cm]{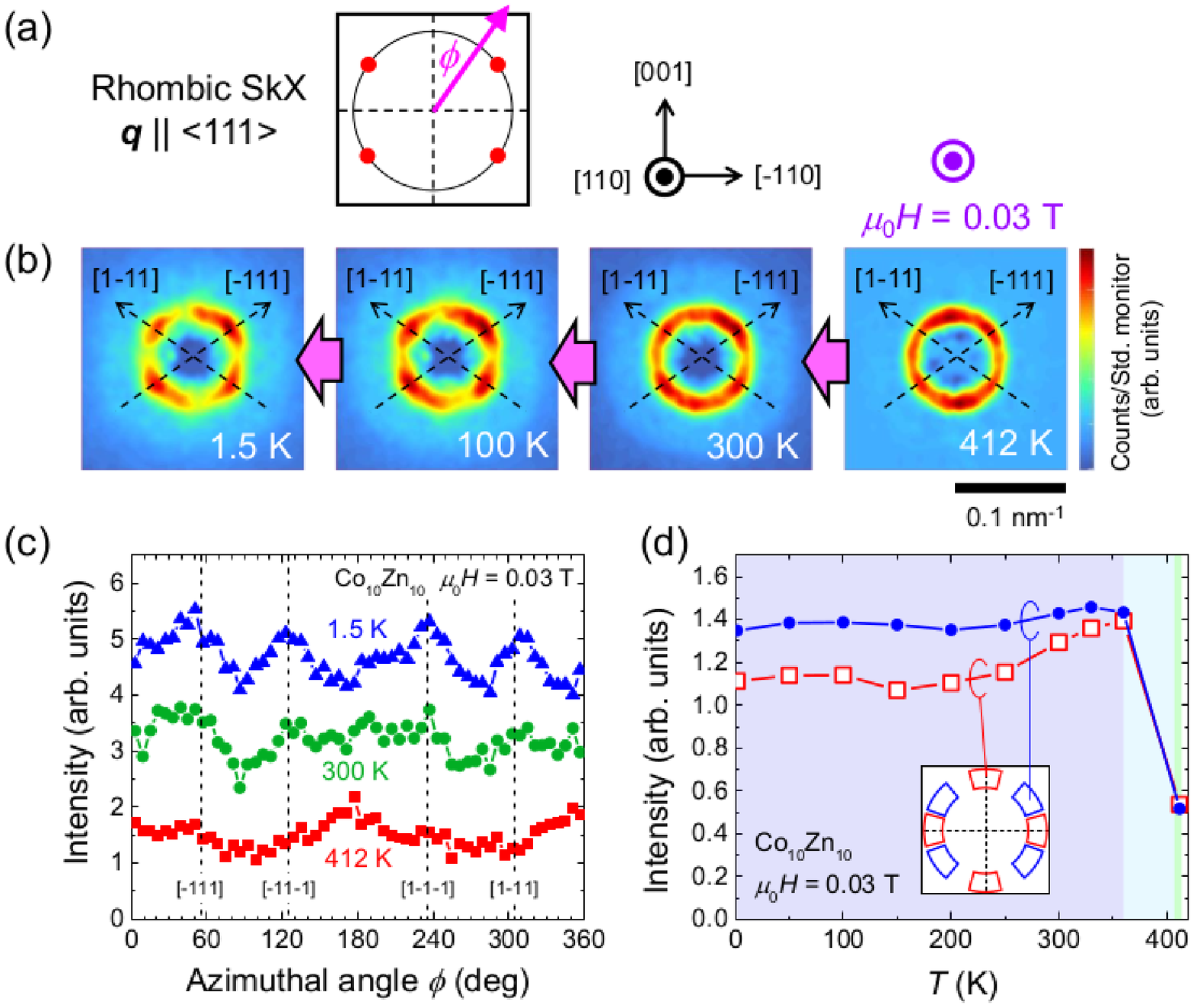}
\end{center}
\end{figure}
%%%%%%%%%%%%%%%%%  FIG6  %%%%%%%%%%%%%%
\noindent
FIG. 6. Metastable SkX state in \ten\ during a field cooling process at 0.03 T. 
(a) Schematic SANS pattern on the (110) plane, perpendicular to the field, expected for a rhombic SkX with $\vect{q}$ $\parallel$ $<$111$>$. 
(b) SANS images observed at 412 K, 300 K, 100 K and 1.5 K during the FC process at 0.03 T. 
The intensity scale of the color plot varies between each panel. The magnetic field was applied at 412 K after ZFC. 
(c) Azimuthal angle ($\phi$) dependence of SANS intensity at 412 K, 300 K and 1.5 K. 
The intensity data for 1.5 K is vertically shifted by 2 for clarity. The angles corresponding to the [-111] and [1-11] directions ($\phi$ = 55$^\circ$, 125$^\circ$, 235$^\circ$ and 305$^\circ$) are indicated by dashed lines. 
(d) Temperature dependence of the SANS intensity. 
Blue symbols denote the intensity integrated over the region close to the [-111] and [1-11] directions ($\phi$ = 55$^\circ$, 125$^\circ$, 235$^\circ$ and 305$^\circ$ with the width of $\Delta\phi$ = 30$^\circ$; blue area in the inset). 
The intensity integrated over the region around the [001] and [-110] directions ($\phi$ = 0$^\circ$, 90$^\circ$, 180$^\circ$ and 270$^\circ$ with the width of $\Delta\phi$ = 30$^\circ$; red area in the inset) are denoted by red symbols.
The temperature regions of the equilibrium triangular SkX, metastable triangular SkX and metastable rhombic SkX are indicated with the light-green, light-blue and purple shadings, respectively.
\\

%\newpage

%%%%%%%%%%%%%%%%%  FIG7  %%%%%%%%%%%%%%
\begin{figure}[htbp]
\begin{center}
\includegraphics[width=16cm]{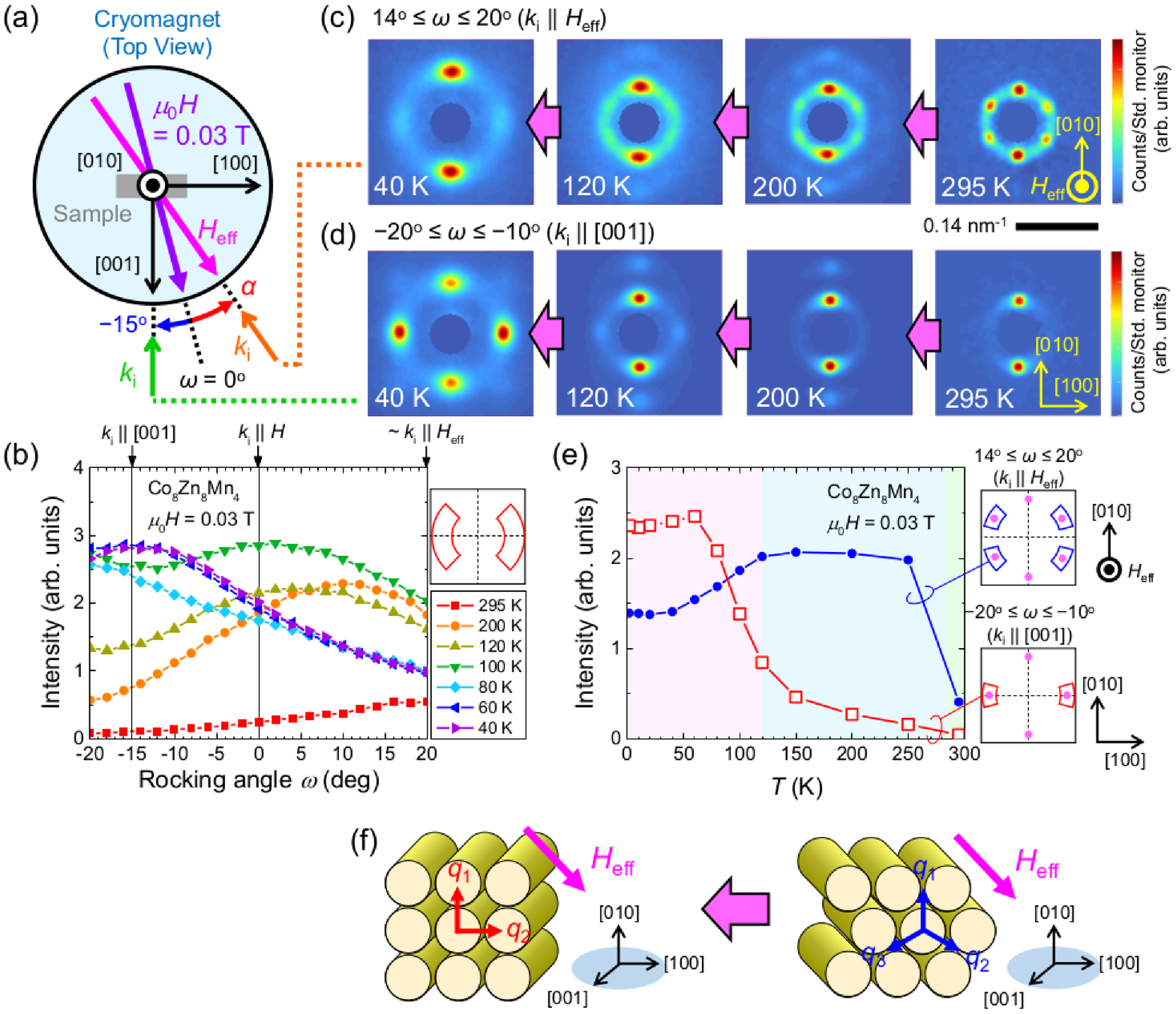}
\end{center}
\end{figure}
%%%%%%%%%%%%%%%%%  FIG7  %%%%%%%%%%%%%%
\noindent
FIG. 7. Reorientation of metastable SkX at low temperatures in \eight\ under an off-axis field. 
(a) Schematic top view of the experimental configuration. 
Rocking angle ($\omega$) is defined as the angle between the incident neutron beam ($k_\mathrm{i}$) and the applied magnetic field ($H$).
In this experiment, magnetic field was tilted by 15$^\circ$ away from the [001] direction.
Due to the demagnetization effect from the tilted magnetic field applied to the rectangular-shaped sample (gray object), the effective magnetic field ($H_\mathrm{eff}$, pink arrow) inside the sample is tilted by more than 15$^\circ$ from [001] direction. 
The additional tilting angle ($\alpha$) of $H_\mathrm{eff}$ from $H$ is estimated to be $\alpha$ $\sim$ 20$^\circ$ according to both the calculation of the demagnetization field\cite{Demag} and the peak position of rocking curve at 295 K shown in panel (b). 
(b) Rocking curves at selected temperatures in the field cooling (FC) process. SANS intensities are integrated over the azimuthal angle area at $\phi$ = 90$^\circ$, 270$^\circ$ with the width of $\Delta\phi$ = 90$^\circ$ (red area in the inset). 
(c, d) SANS images at 295 K, 200 K, 120 K and 40 K during a FC process at 0.03 T. These SANS images are averaged over the limited rocking angles: (c) 14$^\circ$ $\leq$ $\omega$ $\leq$ 20$^\circ$ (around $H_\mathrm{eff}$) and (d) $-$20$^\circ$ $\leq$ $\omega$ $\leq$ $-$10$^\circ$ (around [001]). 
The intensity scale of the color plot varies between each panel. 
(e) Temperature dependence of the SANS intensities during the FC process. Blue closed circles show the SANS intensity integrated over the azimuthal angle area at $\phi$ = 60$^\circ$, 120$^\circ$, 240$^\circ$, 300$^\circ$ with the width of $\Delta\phi$ = 30$^\circ$ in the rocking angle range of 14$^\circ$ $\leq$ $\omega$ $\leq$ 20$^\circ$ (blue area in the inset). Red open squares represent the SANS intensity integrated over the azimuthal angle area at $\phi$ = 90$^\circ$, 180$^\circ$ with the width of $\Delta\phi$ = 30$^\circ$ in the rocking angle range of $-$20$^\circ$ $\leq$ $\omega$ $\leq$ $-$10$^\circ$ (red area in the inset).
The temperature regions of the equilibrium triangular SkX, metastable triangular SkX and metastable square SkX are indicated with the light-green, light-blue and pink shadings, respectively.
(f) Schematic illustration of transformation from a triangular SkX to a square SkX under the off-axis effective field $H_\mathrm{eff}$ tilted from [001]. While the skyrmions in the triangular SkX are parallel to $H_\mathrm{eff}$, those in the square SkX at low temperatures are parallel to [001]. 
\\

\newpage

%%%%%%%%%%%%%%%%%  FIG8  %%%%%%%%%%%%%%
\begin{figure}[htbp]
\begin{center}
\includegraphics[width=14cm]{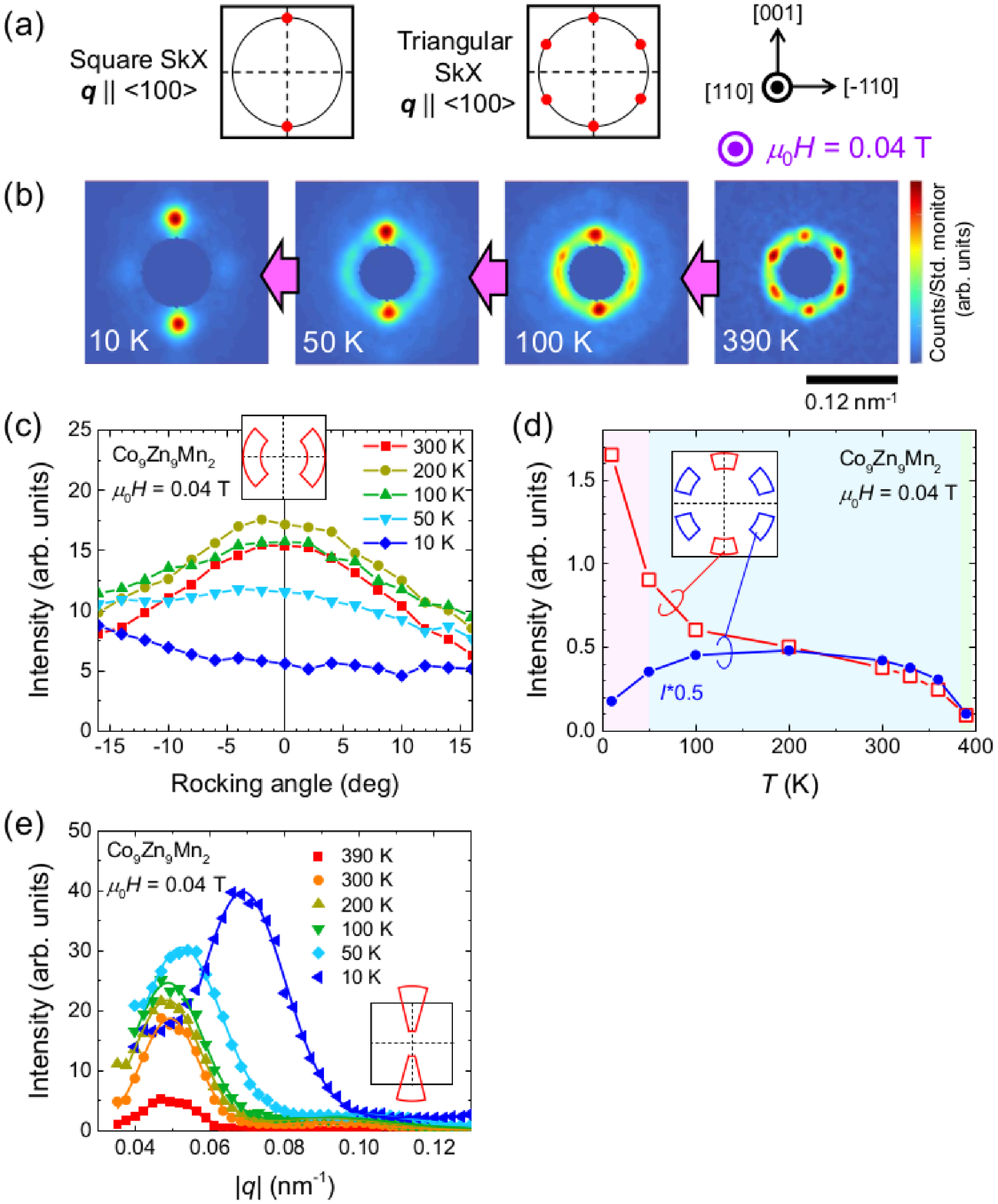}
\end{center}
\end{figure}
%%%%%%%%%%%%%%%%%  FIG8  %%%%%%%%%%%%%%
\noindent
FIG. 8. Metastable SkX state in \nine. (a) Schematic SANS pattern on the (110) plane, perpendicular to the field, expected for a triangular SkX state (right panel) and for a square SkX state (left panel). The triangular SkX state with one of triple-$\vect{q}$ $\parallel$ [001] shows 6 spots at $\phi$ = 0$^\circ$, 60$^\circ$, 120$^\circ$, 180$^\circ$, 240$^\circ$, 300$^\circ$. Here, $\phi$ is defined as the clockwise azimuthal angle from the vertical [001] direction. The square SkX state with double-$\vect{q}$ $\parallel$ [001] and $\parallel$ [100] or [010] (out of the plane) exhibits 2 spots at $\phi$ = 0$^\circ$, 180$^\circ$. 
(b) SANS images observed at 390 K, 100 K, 50 K and 10 K in the field cooling (FC) process at 0.04 T. The intensity scale of the color plot varies between each panel. The magnetic field was applied at 390 K after zero field cooling. 
(c) Rocking curves at selected temperatures in the FC process. SANS intensities are integrated over the azimuthal angle area at $\phi$ = 90$^\circ$, 270$^\circ$ with the width of $\Delta\phi$ = 90$^\circ$ (red area in the inset). The origin of the rocking angle ($\omega$ = 0$^\circ$) corresponds to $k_\mathrm{i}$ $\parallel$ $H$ $\parallel$ [110].
(d) Temperature dependence of the SANS intensities. 
Blue closed circles show the SANS intensity integrated over the azimuthal angle area at $\phi$ = 60$^\circ$, 120$^\circ$, 240$^\circ$, 300$^\circ$ with the width of $\Delta\phi$ = 30$^\circ$  (blue area in the inset) and finally divided by 2. 
Red open squares represent the SANS intensity integrated over the azimuthal angle area at $\phi$ = 0$^\circ$, 180$^\circ$ with the width of $\Delta\phi$ = 30$^\circ$ (red area in the inset).
The temperature regions of the equilibrium triangular SkX, metastable triangular SkX and metastable square SkX are indicated with the light-green, light-blue and pink shadings, respectively.
(e) Radial $|q|$ dependence of SANS intensity, integrated over the azimuthal angle area at $\phi$ = 0$^\circ$, 180$^\circ$ with the width of $\Delta\phi$ = 30$^\circ$ (red area in the inset), at several temperatures during the FC process. The data points are fitted to a Gaussian function (solid line).
\\

\newpage

%%%%%%%%%%%%%%%%%  FIG9  %%%%%%%%%%%%%%
\begin{figure}[htbp]
\begin{center}
\includegraphics[width=14cm]{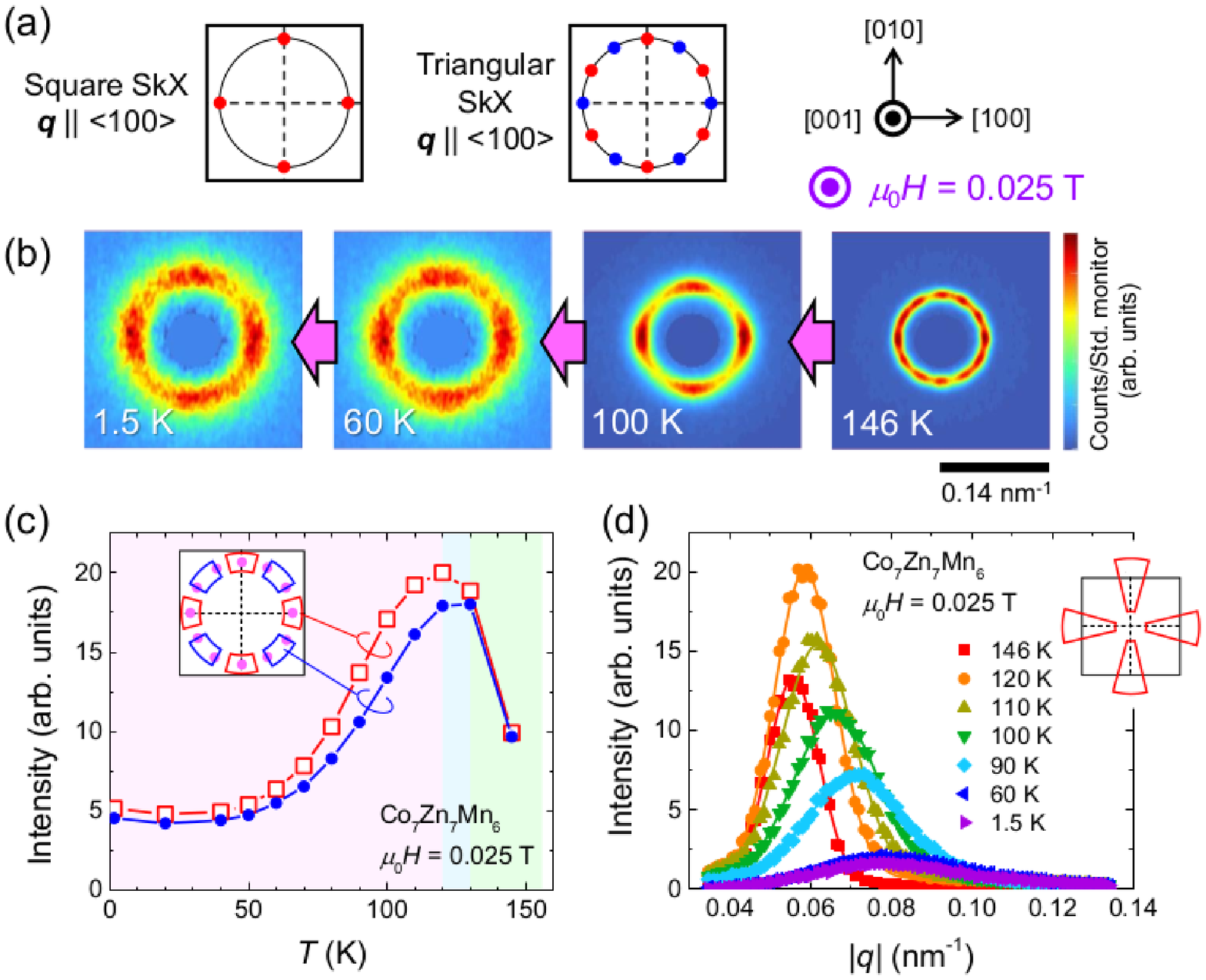}
\end{center}
\end{figure}
%%%%%%%%%%%%%%%%%  FIG9  %%%%%%%%%%%%%%
\noindent
FIG. 9. Temperature dependence of metastable SkX state in \seven. 
(a) Schematic SANS pattern on the (001) plane, perpendicular to the field, expected for a triangular SkX state (right panel) and for a square SkX state (left panel). The triangular SkX state forms two domains with one of the triple-$\vect{q}$ $\parallel$ [100] (6 blue spots) and  $\parallel$ [010] (6 red spots), respectively, resulting in 12 spots at every 30$^\circ$ from $\phi$ = 0$^\circ$. Here, $\phi$ is defined as the clockwise azimuthal angle from the vertical [010] direction. The square SkX state with double-$\vect{q}$ $\parallel$ [100] and [010] shows 4 spots at $\phi$ = 0$^\circ$, 90$^\circ$, 180$^\circ$, 270$^\circ$.
(b) SANS images observed at 146 K, 100 K, 60 K and 1.5 K during the field cooling (FC) process at 0.025 T. 
The intensity scale of the color plot varies between each panel. The magnetic field was applied at 146 K after zero field cooling. 
(c) Temperature dependence of the SANS intensities. Blue closed circles show the SANS intensity integrated over the azimuthal angle area at $\phi$ = 45$^\circ$, 135$^\circ$, 225$^\circ$, 315$^\circ$ with the width of $\Delta\phi$ = 30$^\circ$ (blue area in the inset). 
Red open squares represent the SANS intensity integrated over the azimuthal angle area at $\phi$ = 0$^\circ$, 90$^\circ$, 180$^\circ$, 270$^\circ$ with the width of $\Delta\phi$ = 30$^\circ$ (red area in the inset). 
The temperature regions of the equilibrium triangular SkX, metastable triangular SkX and metastable square SkX are indicated with the light-green, light-blue and pink shadings, respectively.
(d) Radial $|q|$ dependence of SANS intensity, integrated over the azimuthal angle area at $\phi$ = 0$^\circ$, 90$^\circ$, 180$^\circ$, 270$^\circ$ with the width of $\Delta\phi$ = 30$^\circ$ (red area in the inset), at several temperatures in the FC process. The data points are fitted to a Gaussian function (solid line).
\\

\newpage

%%%%%%%%%%%%%%%%%  FIG10  %%%%%%%%%%%%%%
\begin{figure}[htbp]
\begin{center}
\includegraphics[width=16cm]{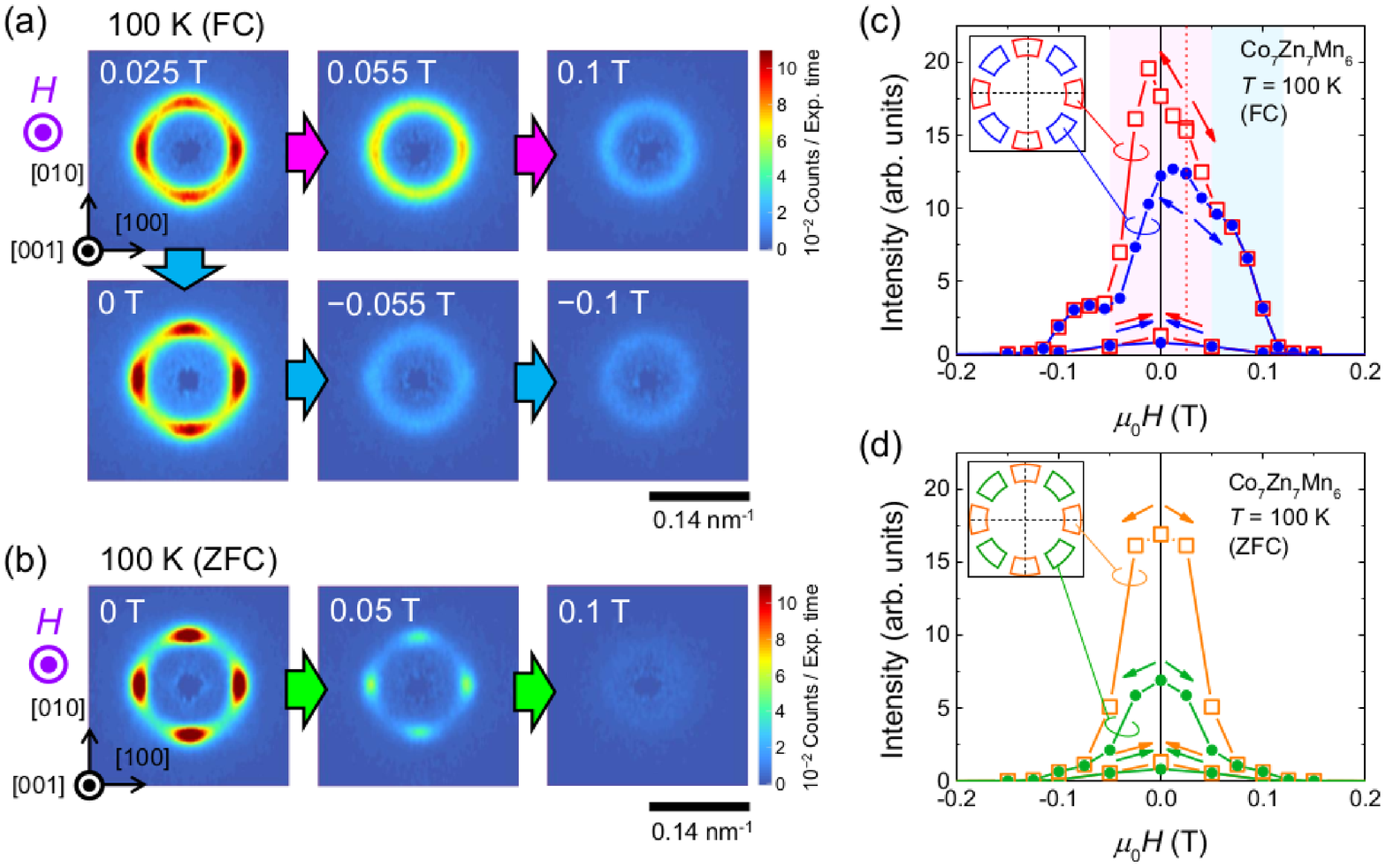}
\end{center}
\end{figure}
%%%%%%%%%%%%%%%%%  FIG10  %%%%%%%%%%%%%%
\noindent
FIG. 10. Field effect on metastable SkX state and helical state in \seven\ at 100 K. 
(a) SANS images at selected fields during field scans to the positive and negative directions at 100 K after a field cooling (FC) at 0.025 T. 
(b) SANS images observed at 0 T, 0.05 T and 0.1 T at 100 K after a zero-field cooling (ZFC) (reproduced from our previous work in Ref. \cite{Karube_776}. Copyright 2018, American Association for the Advancement of Science). 
The intensity scale of the color plot is fixed for all the panels in (a) and (b).
(c) Field dependence of the SANS intensities at 100 K after the FC defined similarly as in Fig. 9(c). 
%Blue closed circles show the SANS intensity integrated over the azimuthal angle area at $\phi$ = 45$^\circ$, 135$^\circ$, 225$^\circ$, 315$^\circ$ with the width of $\Delta\phi$ = 30$^\circ$ (blue area in the inset). 
%Red open squares represent the SANS intensity integrated over the azimuthal angle area at $\phi$ = 0$^\circ$, 90$^\circ$, 180$^\circ$, 270$^\circ$ with the width of $\Delta\phi$ = 30$^\circ$ (red area in the inset). 
%Here, $\phi$ is defined as the clockwise azimuthal angle from the vertical [010] direction.
The field regions of the metastable triangular SkX and the metastable square SkX are indicated with the light-blue and pink shadings, respectively. 
(d) Field dependence of the SANS intensities at 100 K after the ZFC (reproduced from our previous work in Ref. \cite{Karube_776}. Copyright 2018, American Association for the Advancement of Science). 
Green closed circles show the SANS intensity integrated over the azimuthal angle area at $\phi$ = 45$^\circ$, 135$^\circ$, 225$^\circ$, 315$^\circ$ with the width of $\Delta\phi$ = 30$^\circ$ (green area in the inset). 
Orange open squares represent the SANS intensity integrated over the azimuthal angle area at $\phi$ = 0$^\circ$, 90$^\circ$, 180$^\circ$, 270$^\circ$ with the width of $\Delta\phi$ = 30$^\circ$ (orange area in the inset). 
\\

\newpage

%%%%%%%%%%%%%%%%%  FIG11  %%%%%%%%%%%%%%
\begin{figure}[htbp]
\begin{center}
\includegraphics[width=16cm]{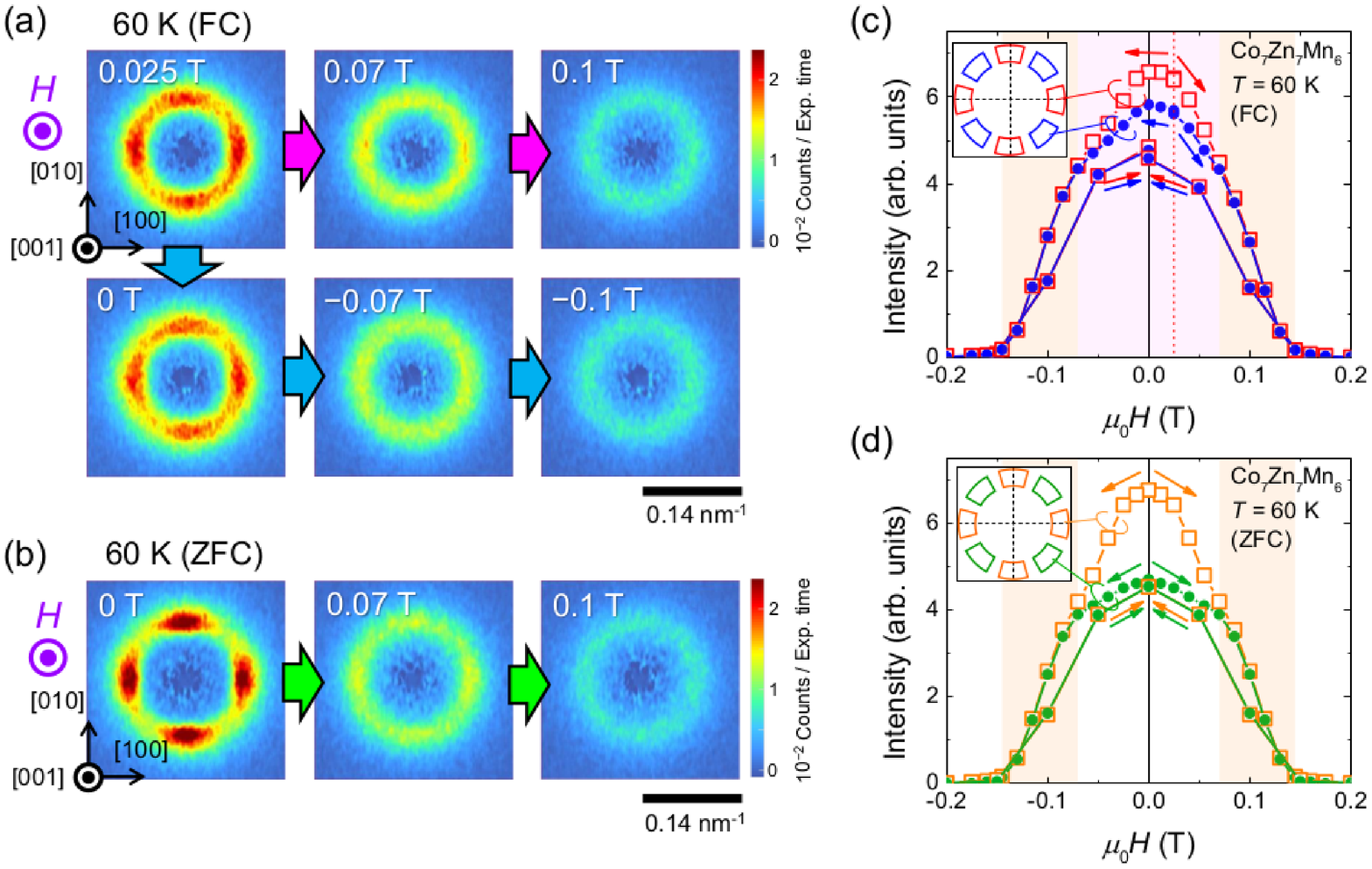}
\end{center}
\end{figure}
%%%%%%%%%%%%%%%%%  FIG11  %%%%%%%%%%%%%%
\noindent
FIG. 11. Metastable and low-temperature equilibrium skyrmion states in magnetic fields in \seven\ at 60 K. 
(a) SANS images at selected fields during field scans to the positive and negative directions at 60 K after a FC at 0.025 T. 
(b) SANS images observed at 0 T, 0.07 T and 0.1 T at 60 K after a ZFC. 
The intensity scale of the color plot is fixed for all the panels in (a) and (b).
(c) Field dependence of the SANS intensities at 60 K after the FC, integrated over the azimuthal angle areas defined similarly as in Fig. 9(c). 
The field regions of the metastable square SkX and the equilibrium disordered skyrmions (DSk) are indicated with the pink and orange shadings, respectively. 
(d) Field dependence of the SANS intensities at 60 K after the ZFC (reproduced from our previous work in Ref. \cite{Karube_776}. Copyright 2018, American Association for the Advancement of Science), integrated over the azimuthal angle areas defined similarly as in Fig. 10(d).  
The field region of the equilibrium DSk phase is indicated with the orange shading. 
\\

\newpage

%%%%%%%%%%%%%%%%%  FIG12  %%%%%%%%%%%%%%
\begin{figure}[htbp]
\begin{center}
\includegraphics[width=14cm]{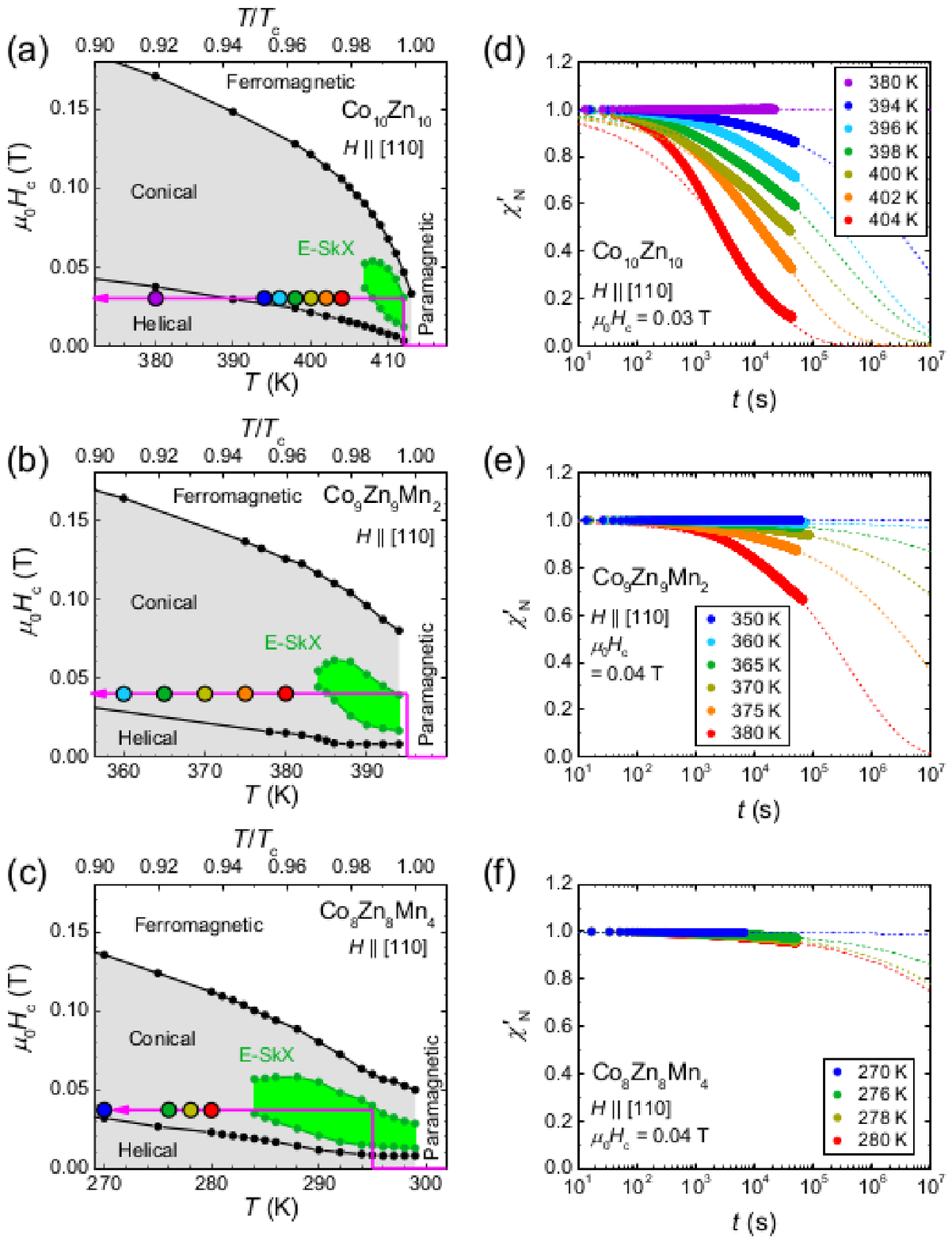}
\end{center}
\end{figure}
%%%%%%%%%%%%%%%%%  FIG12  %%%%%%%%%%%%%%
\noindent
FIG. 12. (a-c) Temperature ($T$) - magnetic field ($H$) phase diagrams near $T_\mathrm{c}$ in (a) \ten, (b) \nine\ and (c) \eight, determined by ac susceptibility ($\chi^\prime$) measurements with $H$ $\parallel$ [110]. 
For the magnetic field, calibrated values $H_\mathrm{c}$ are used.
The displayed range of $T/T_\mathrm{c}$ (upper horizontal axis) is fixed to be 0.90 - 1.01 for the three panels. 
Time-dependent $\chi^\prime$ measurements after a field cooling (FC, pink arrow) via the equilibrium SkX phase (green region) are performed at several temperatures denoted with circles, whose color corresponds to the color of data points in panels (d-f).
(d-f) Time dependence of the normalized ac susceptibility, defined as $\chi^\prime_\mathrm{N}(t) \equiv [\chi^\prime (\infty) - \chi^\prime (t)]/[\chi^\prime (\infty) - \chi^\prime (0)]$, for (d) \ten, (e) \nine\ (reproduced from our previous work in Ref. \cite{Karube_992} with permissions. Copyright 2017, American Physical Society) and (f) \eight, measured in the processes described in panels (a-c). 
Here, $\chi^\prime (0)$ is an initial value (metastable SkX state), and $\chi^\prime (\infty)$ is the value for the equilibrium conical state which, as a fully relaxed state is assumed to be the value of $\chi^\prime$ at the same magnitude of field after a field decreasing run from a ferromagnetic phase.
The data points are fitted to stretched exponential functions (dotted lines), $\exp \left\{ -(t/\tau)^\beta \right\}$. 
\\

\newpage

%%%%%%%%%%%%%%%%%  FIG13  %%%%%%%%%%%%%%
\begin{figure}[htbp]
\begin{center}
\includegraphics[width=8cm]{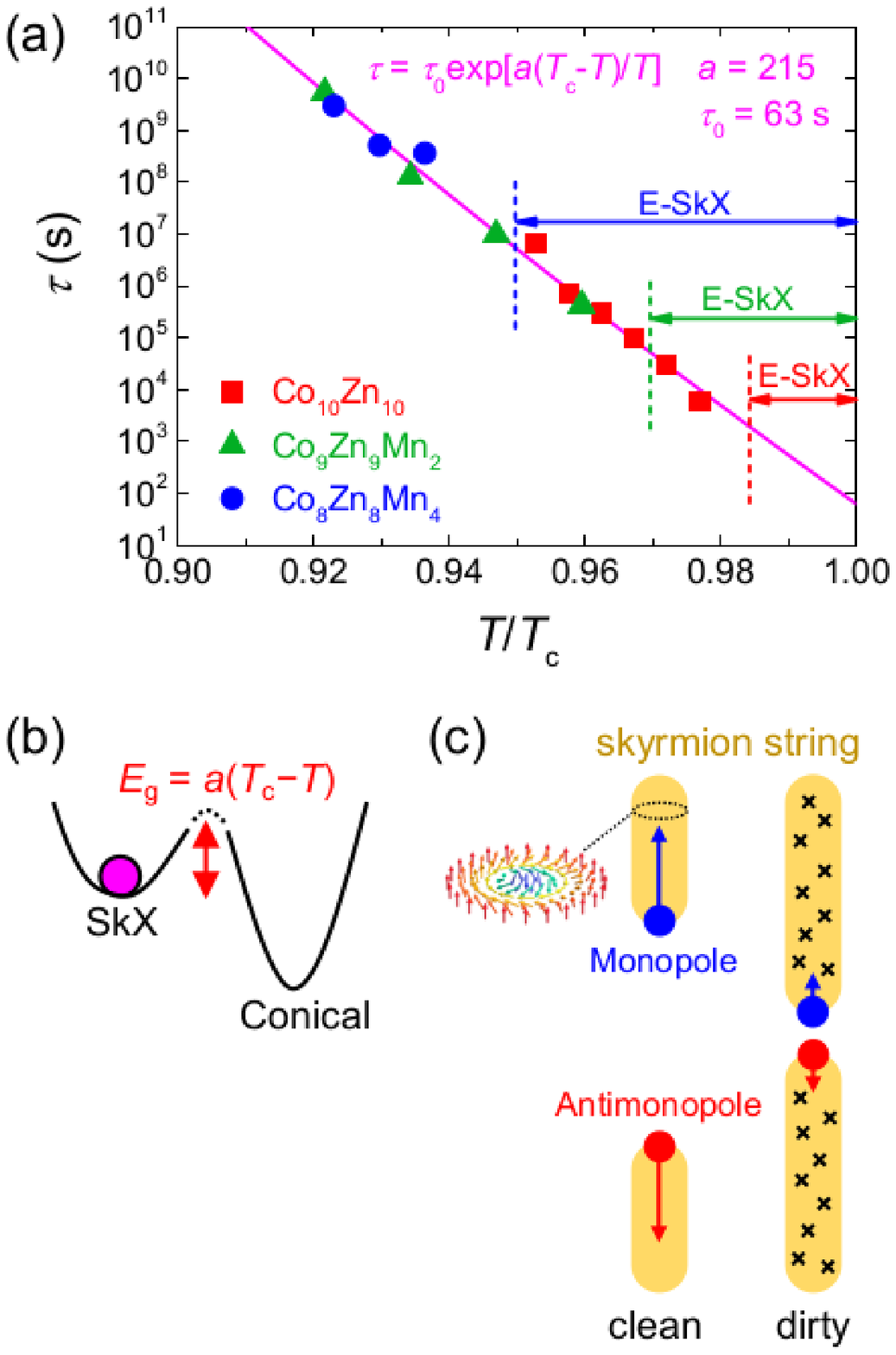}
\end{center}
\end{figure}
%%%%%%%%%%%%%%%%%  FIG13  %%%%%%%%%%%%%%
\noindent
FIG. 13. (a) Relaxation time ($\tau$) of metastable SkX states plotted against normalized temperature $T/T_\mathrm{c}$ in \ten\ (red squares), \nine\ (green triangles, reproduced from our previous work in Ref. \cite{Karube_992} with permissions. Copyright 2017, American Physical Society) and \eight\ (blue circles). The $\tau$ values are determined by the fits in Figs. 12(d-f).
The solid arrows and dotted lines indicate temperature ranges and lower boundaries of equilibrium SkX phases, respectively.
The data points from the three different compositions, showing a good scaling, are fitted to a modified Arrhenius law (pink solid line), $\tau = \tau_0 \exp \left\{ a(T_\mathrm{c} - T)/T\right\}$, with an assumption of a temperature-dependent activation energy $E_\mathrm{g} = a(T_\mathrm{c}-T)$ as described in Ref. \cite{Oike,Karube_992,Birch}. The obtained parameters are $a$ = 215 and $\tau_0$ = 63 s.
(b) Schematic illustration of free energy landscape with a metastable SkX state and the most stable conical state. (c) Schematic illustration of the destruction process for a skyrmion string induced by movement of an emergent monopole-antimonopole pair for a clean system and a dirty system. 
In the dirty system, the propagation of the monopole and antimonopole are hindered by magnetic disorder while less so in the clean system.
\\

%%%%%%%%%%%%%%%%%% TABLE2  %%%%%%%%%%%%%%
\begin{table}[htbp]
\begin{center}
\begin{tabular}{c|ccc} \hline
\:\:\: Material \:\:\: & \:\:\: $a$ \:\:\: & \:\:\: $\tau_0$ (s) \:\:\: & \:\:\: Reference \:\:\: \\ \hline
\:\:\: Co-Zn-Mn \:\:\: & \:\:\: 215 \:\:\: & \:\:\: 63 \:\:\: & \:\:\: This work \:\:\: \\ 
\:\:\: MnSi \:\:\: & \:\:\: 65 \:\:\: & \:\:\: 2.7 $\times$ 10$^{-4}$ \:\:\: & \:\:\: \cite{Oike} \:\:\: \\ 
\:\:\: Cu$_2$OSeO$_3$ \:\:\: & \:\:\: 96 \:\:\: & \:\:\: 3 \:\:\: & \:\:\: \cite{Birch} \:\:\: \\ 
Cu$_2$OSeO$_3$ (Zn 2.5\% doped) & \:\:\: 94 \:\:\: & \:\:\: 150 \:\:\: & \:\:\: \cite{Birch} \:\:\: \\ \hline
\end{tabular}
\end{center}
\end{table}
%%%%%%%%%%%%%%%%%%  TABLE2  %%%%%%%%%%%%%%
\noindent
TABLE. 2. Summary of relaxation parameters for metastable SkX in various materials. 
\\

\end{document}

\title{Supplementary materials for \\ ``Metastable skyrmion lattices governed by magnetic disorder and anisotropy in $\beta$-Mn-type chiral magnets"}
\author{K.~Karube}
\altaffiliation{These authors equally contributed to this work}
\affiliation{RIKEN Center for Emergent Matter Science (CEMS), Wako 351-0198, Japan.}
\author{J.~S.~White}
\altaffiliation{These authors equally contributed to this work}
\affiliation{Laboratory for Neutron Scattering and Imaging (LNS), Paul Scherrer Institute (PSI),
CH-5232 Villigen, Switzerland.}
\author{V.~Ukleev}
\affiliation{Laboratory for Neutron Scattering and Imaging (LNS), Paul Scherrer Institute (PSI),
CH-5232 Villigen, Switzerland.}
\author{C.~D.~Dewhurst}
\affiliation{Institut Laue-Langevin (ILL), 71 avenue des Martyrs, CS 20156, 38042 Grenoble cedex 9, France}
\author{R.~Cubitt}
\affiliation{Institut Laue-Langevin (ILL), 71 avenue des Martyrs, CS 20156, 38042 Grenoble cedex 9, France}
\author{A.~Kikkawa}
\affiliation{RIKEN Center for Emergent Matter Science (CEMS), Wako 351-0198, Japan.}
\author{Y.~Tokunaga}
\affiliation{Department of Advanced Materials Science, University of Tokyo, Kashiwa 277-8561, Japan.}
\author{H.~M.~R\o nnow}
\affiliation{Laboratory for Quantum Magnetism (LQM), Institute of Physics, \'Ecole Polytechnique F\'ed\'erale de Lausanne (EPFL), CH-1015 Lausanne,
Switzerland.}
\author{Y.~Tokura}
\affiliation{RIKEN Center for Emergent Matter Science (CEMS), Wako 351-0198, Japan.}
\affiliation{Department of Applied Physics and Tokyo College, University of Tokyo, Bunkyo-ku 113-8656, Japan.}
\author{Y.~Taguchi}
\affiliation{RIKEN Center for Emergent Matter Science (CEMS), Wako 351-0198, Japan.}

\maketitle

\newpage
\subsection{Magnetization curve at 2 K}
In Fig. S1(a), we show field dependence of magnetization at 2 K measured up to 7 T in \ten, \nine, \eight\ and \seven.
The saturation magnetization ($M_\mathrm{s}$), which is defined as the value at 2 K and 7 T and summarized in Table 1 in the main text, takes a maximum at \nine\ ($M_\mathrm{s}$ = 14.3 $\mu_\mathrm{B}$/f.u.), while the helimagnetic transition temperature \Tc\ monotonically decreases from \ten\ as the Mn concentration is increased. 
This indicates that Co spin and Mn spins are ferromagnetically coupled at least in the low Mn concentration region.
As seen in the magnified view around the low-field region [Fig. S1(b)], hysteresis is observed for $-$0.3 T $\leq$ $\mu_0 H$ $\leq$ 0.3 T for \eight\ and for $-$0.7 T $\leq$ $\mu_0 H$ $\leq$ 0.7 T for \seven.
The hysteresis is observed only below the reentrant spin glass transition temperature $T_\mathrm{g}$ in \eight\ ($T_\mathrm{g}$ $\sim$ 10 K) and \seven\ ($T_\mathrm{g}$ $\sim$ 30 K). 
The hysteresis is not discernible in \ten\ and \nine, neither of which exhibits the reentrant spin glass transition down to 2 K.
Similar results of magnetization have been reported in Ref. \cite{Bocarsly}.

\subsection{Field dependence of metastable SkX at 300 K in \ten}
Here, we show the detailed results of field variation in metastable skyrmion crystal (SkX) at 300 K in \ten\ (Fig. S2).

Small-angle neutron scattering (SANS) patterns for several magnetic fields at 300 K after a field cooling (FC) with 0.03 T are shown in Fig. S2(a). 
In the field sweeping process toward positive direction (pink arrows), the SANS pattern with stronger intensity along the [-111] and [1-11] direction changes to a uniform ring at 0.1 T similar to that observed in the equilibrium SkX phase at 412 K at 0.03 T. This field variation corresponds to the change in lattice form of skyrmions from rhombic-like one to the original triangular one. The ring-like SANS signal corresponding to the triangular SkX persists up to 0.2 T. 
In the field sweeping run toward negative direction (light-blue arrows) down to $-$0.06 T, the intensity along the [-111] and [1-11] directions increases and clear 4 spots are observed, indicating that lattice form of skyrmions approaches ideal rhombic one.
Upon further increasing the field in the negative direction, the SANS intensity from the rhombic SkX disappears at $-$0.1 T.
For comparison, SANS patterns at the same magnetic fields at 300 K after a zero-field cooling (ZFC) are presented in Fig. S2(b).
The SANS pattern with broad intensity around the [1-11] direction originating from a helical state disappears at 0.1 T, which is totally distinct from the field variation in the metastable SkX after the FC.

The real-part of ac susceptibility ($\chi^\prime$) and the SANS intensities at 300 K after the FC are plotted against applied field in Fig. S2(c) and S2(d), respectively. 
Here, ZFC data are also plotted in the same graph for comparison.
The small values of $\chi^\prime$ and the large SANS intensities observed for $-$0.1 T $\leq$ $\mu_0 H$ $\leq$ 0.2 T in the initial field scans, as compared to those in the returning runs, are attributed to the metastable SkX state. 
Thus, the metastable SkX state survives over a wide field region up to the conical-ferromagnetic phase boundary and down to the negative fields as marked with blue triangles in Fig. S2(c).
The boundary between the rhombic SkX and the triangular SkX is determined as the magnetic field where the SANS intensity around the [-111] and [1-11] directions becomes higher than that around the [001] and [-110] directions [purple triangle in Fig. S2(d)], and plotted in the state diagram of Fig. 2(b) in the main text. 

\subsection{Field dependence of metastable SkX at 1.5 K in \ten}
We also discuss field variation in metastable SkX in \ten\ at 1.5 K (Fig. S3).

Figure S3(a) shows several SANS patterns at 1.5 K for selected fields after the FC with 0.03 T. 
In the field sweeping toward positive direction, the broad 4-spot pattern originating from the rhombic lattice of skyrmions persists up to 0.10 T and finally transforms to a ring-like pattern with strong intensity at 0.16 T. 
In the field sweeping process toward negative direction, the 4-spot pattern is preserved down to $-$0.16 T before Bragg spots disappear at $-$0.20 T.
This result demonstrates that the metastable rhombic SkX survives over a wider field region than that observed at 300 K but finally original triangular lattice is restored at high magnetic fields. 
For comparison we also show field variation in the SANS pattern at 1.5 K after ZFC in Fig. S3(b).
The 4 spots along the [-111] and [1-11] direction, corresponding to the helical multi-domain state with $\vect{q}$ $\parallel$ $<$111$>$, are observed up to 0.16 T with gradual decreases in scattering intensity and $q$ position. 
However, a ring-like patten is never observed at any fields, which is distinct from the result after the FC.

We show the field dependence of $\chi^\prime$ at 5 K and SANS intensities at 1.5 K after the FC in Fig. S3(c) and S3(d), respectively. 
The both measurements exhibit large and asymmetric hysteresis, corresponding to the metastable SkX state as indicated with blue triangles.
A boundary of rhombic-triangular skyrmion lattices [purple triangle in Fig. S3(d)] is determined similarly to the previous section. 
The field dependence of $\chi^\prime$ is complex as compared with the result at 300 K [Fig. S2(c)], and the hysteresis subsists to $-$0.3 T in the negative field region, while the SANS intensity originating from the metastable SkX state disappears at $-$0.2 T.
These results indicate that a helical state with $\vect{q}$ vector along the [11-1] or [111] direction, which is out of the (110) plane and thus not detectable in the present SANS configuration, persists up to the field-induced ferromagnetic phase boundary after complete destruction of the metastable SkX state around $-$0.2 T.
For comparison, the field dependence of $\chi^\prime$ at 5 K and SANS intensities at 1.5 K after ZFC are also displayed in the same figures. 
In this case, a hysteresis in $\chi^\prime$ corresponding to the helical multi-domain state is observed between $-$0.3 T and 0.3 T, which is wider than the hysteresis region (between $-$0.2 T and 0.2 T) in the SANS intensity that represents the contribution from helical states with $\vect{q}$ $\parallel$ [-111] and [1-11].
These ZFC results also indicate that a helical state with $\vect{q}$ $\parallel$ [11-1] and [111] remains above 0.2 T.
The complex field dependence is attributed to the strong magnetocrystalline anisotropy that favors $\vect{q}$ $\parallel$ $<$111$>$.
As described in the next section, the helical state forms a chiral soliton lattice under magnetic fields due to the strong anisotropy of $\vect{q}$ vector.
In the field returning process from high fields both after the FC and ZFC, a clear dip structure in $\chi^\prime$ and relatively large SANS intensity are observed between $-$0.1 T and 0.1 T. 
This is probably because helical $\vect{q}$ vector flops from the field direction to within the plane perpendicular to the field.

\subsection{Chiral soliton lattice in \ten}
In this section, we describe possible formation of chiral soliton lattice in \ten\ [Fig. S4].
As presented in Fig. S3(b, d), the 4-spot SANS pattern, which corresponds to the 2-domain helical state with $\vect{q}$ $\parallel$ [-111] and [1-11], persists up to high fields near the ferromagnetic region.
In this field variation, weak higher-harmonic scattering signals are observed along the [1-11] direction away from the strong main spots, which is more clearly discerned in the SANS image at 0.02 T plotted in logarithmic scale in Fig. S4(a).
For quantitative discussion, the radial $q$ dependence of SANS intensity along the [1-11] direction is plotted in Fig. S4(b).
At zero field, the Bragg peak is observed at $q_0$ $\sim$ 0.044 nm$^{-1}$.
As the field is increased to 0.02 T, scattering intensity above 0.07 nm$^{-1}$ is enhanced, and the second, third, and even fourth harmonic peaks ($q_2$, $q_3$ and $q_4$) are clearly observed in addition to the first main peak ($q_1$).
The higher harmonic peaks are still discerned at 0.16 T while the peak positions shift to lower $q$ region.
These peaks are fitted to Gaussian functions and the peak center of $q_n$ ($n$ = 1, 2, 3) normalized by $n q_0$ is plotted against field in Fig. S4(c).
The second and third harmonic peak positions show a good scaling with the first peak position as expected as $q_n$ = $n q_1$.
With increasing field, the value of $q_n/n q_0$ decreases down to 0.8 before SANS signals disappear at 0.24 T.

These results are reminiscent of a chiral soliton lattice (CSL): one-dimensional chain of nonlinear helimagnetic structures as schematically illustrated in Fig. S4(d), which has been observed in monoaxial chiral magnets as represented by CrNb$_3$S$_6$\cite{Togawa}.
Following the theoretical analysis by Togawa \textit{et al}.\cite{Togawa}, field variation in $q$ value for the CSL is expressed as $q(H)/q(0) = \pi^2/[4K(\kappa)E(\kappa)]$, where $K(\kappa)$ and $E(\kappa)$ are the elliptic integrals of the first and second kinds. Here, the elliptic modulus $\kappa$ (0 $<$ $\kappa$ $<$ 1) is determined by energy minimization and given by $\kappa/E(\kappa) = \sqrt{H/H_\mathrm{c}}$, where $H_\mathrm{c}$ is the ferromagnetic saturation field.
This theoretical result for the CSL is plotted with pink dashed line in Fig. S4(c). Here, we set $\mu_0 H_\mathrm{c}$ = 0.24 T, above which the SANS intensity disappears, and found that the observed field variation in $q_n / n q_0$ agrees with the theoretical curve.

A CSL is stabilized under magnetic fields rather than a conventional conical state if $\vect{q}$ vector is strongly anisotropic and thus perpendicular to the fields. The CSL has been also reported in a cubic chiral magnet Cu$_2$OSeO$_3$ under a uniaxial strain which fixes the $\vect{q}$ vector along the strain direction\cite{Okamura}.
In the present study for a cubic chiral magnet \ten\ without uniaxial strain, the CSL behavior is observed probably because $\vect{q}$ vector anisotropy along the $<$111$>$ direction becomes strong at low temperatures, to which the magnetic field is applied perpendicularly. 

\subsection{Transformation of metastable skyrmion lattice in \eight}
Here, we review the metastable SkX state induced by FC and structural transition of the skyrmion lattice within the metastable state in \eight, which was reported in our previous paper\cite{Karube_884} (Fig. S5).
Selected SANS patterns in a FC at 0.04 T from 295 K to 40 K are shown in Fig. S5(b). 
Here, the magnetic field and the incident neutron beam are parallel to the [001] direction.
At 295 K and 0.04 T, inside the equilibrium SkX phase, clear 12 spots are observed.
This 12-spot pattern corresponds to a 2-domain triangular SkX state, in which one of triple-$\vect{q}$ is parallel to the [010] or [100] direction as illustrated in the right panel of Fig. S5(a). 
The triangular SkX persists down to 200 K as a metastable state during the FC.
From 120 K to 40 K, the 12-spot pattern gradually transforms to a 4-spot pattern.
This 4-spot pattern corresponds to a square SkX state with double-$\vect{q}$ vectors parallel to the [010] and [100] directions as illustrated in the left panel of Fig. S5(a). 
In the subsequent re-warming process, the triangular SkX revives already at 200 K.
This reversibility of the metastable SkX can rule out the possibility that the 4-spot pattern is due to a helical multi-domain state with zero topological charge. 

SANS intensity is plotted as a function of temperature in Fig. S5(c). 
Above 120 K, the intensities for the $<$100$>$ and $<$110$>$ directions show similar values as expected for a 12-spot pattern.
Below 120 K, the intensity ratio of $<$100$>$ to $<$110$>$ becomes larger due to the gradual transformation to 4 spots. 
The SANS intensities show maximum at 250 K and decrease below 200 K, indicating evolution of magnetic disorder in the metastable SkX, especially in the square SkX. 

The radial $q$ dependence of the SANS intensity for $<$100$>$ is shown in Fig. S5(d). 
From 120 K to 40 K, where the triangular-square SkX transition occurs, the peak center shows a large shift from 0.055 nm$^{-1}$ to 0.09 nm$^{-1}$ and the peak intensity decreases. 
The temperature dependence of $q$ in the metastable skyrmion state during the field cooling is similar to that of the helical state during zero-field cooling as shown in Fig. 3(g) in the main text.

\subsection{Field dependence of metastable SkX in \nine}
In this section, we describe field variation in the metastable SkX state at low temperatures in \nine\ (Fig. S6).
Selected SANS patterns at 10 K after the FC (0.04 T) are shown in Fig. S6(a). 
The 2-spot pattern at 0.04 T corresponds to the square SkX as discussed in Fig. 8 in the main text.
With increasing field up to 0.3 T, the 2-spot pattern gradually changes to a ring-like one.
The detailed field dependence of the SANS intensity is presented in Fig. S6(c). 
While the large intensity from the 2 vertical spots is rapidly suppressed with increasing field, the weak intensity from the side region hardly decreases, and eventually the intensities from the two regions become similar at 0.3 T. 

As shown in Fig. S6(b), the rocking curve  at 0.3 T clearly shows a peak structure around $\omega$ = 0$^\circ$.
This indicates that the scattering intensity from the ring-like pattern lies in the (110) plane, namely $\vect{q}$ vector is perpendicular to the field, similar to the observation at high temperatures.
Thus, the field variation in the SANS pattern is interpreted as the change from the square SkX with $\vect{q}$ $\parallel$ $<$100$>$ to an orientationally disordered triangular SkX with $\vect{q}$ $\perp$ $H$. 

\newpage
\bibliographystyle{apsrev}

\newpage

%%%%%%%%%%%%%%%%%  FIGS1  %%%%%%%%%%%%%%
\begin{figure}[htbp]
\begin{center}
\includegraphics[width=14cm]{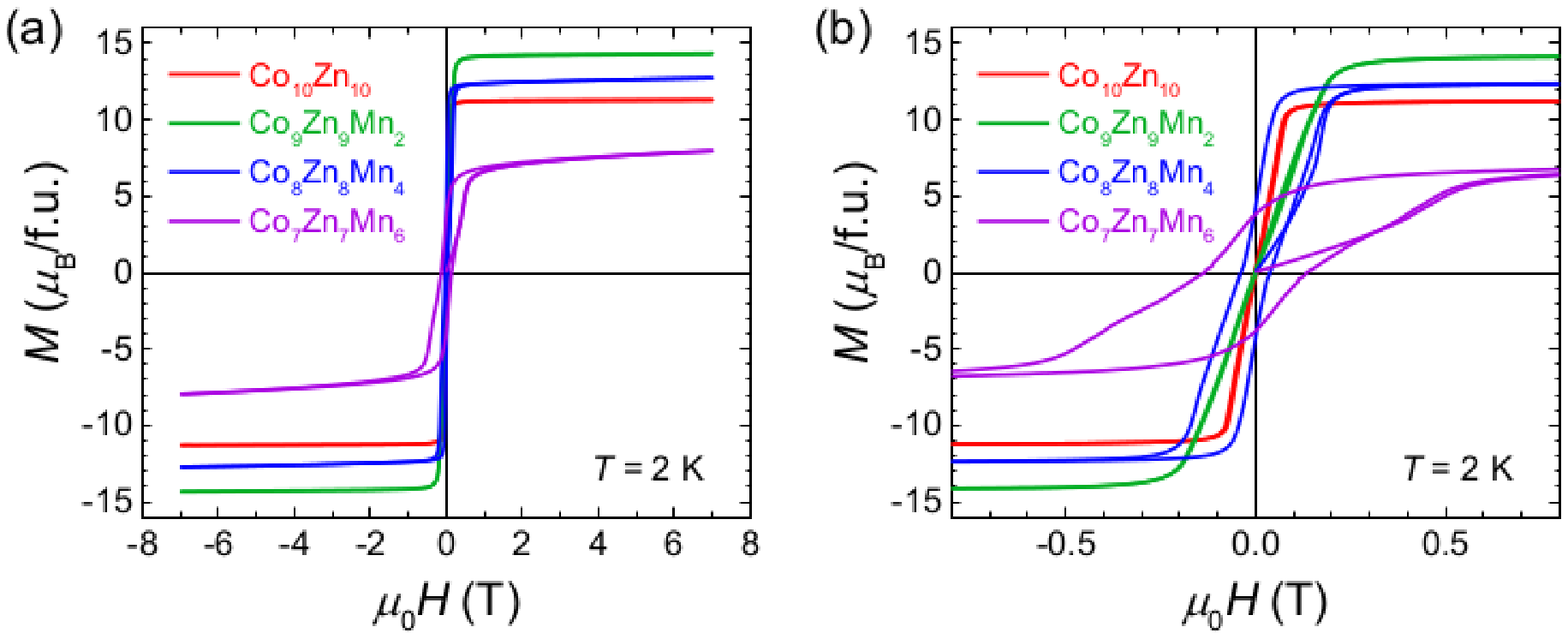}
\end{center}
\end{figure}
%%%%%%%%%%%%%%%%%  FIGS1  %%%%%%%%%%%%%%
\noindent
FIG. S1. (a) Magnetic field ($H$) dependence of magnetization ($M$) at 2 K in single-crystalline samples of \ten, \nine, \eight\ and \seven. 
Magnetic fields are applied along the [110] direction for \ten\ and \nine, and along the [100] direction for \eight\ and \seven, respectively. 
(b) Magnified view of low field region in panel (a). 
\\

\newpage

%%%%%%%%%%%%%%%%%  FIGS2  %%%%%%%%%%%%%%
\begin{figure}[htbp]
\begin{center}
\includegraphics[width=16cm]{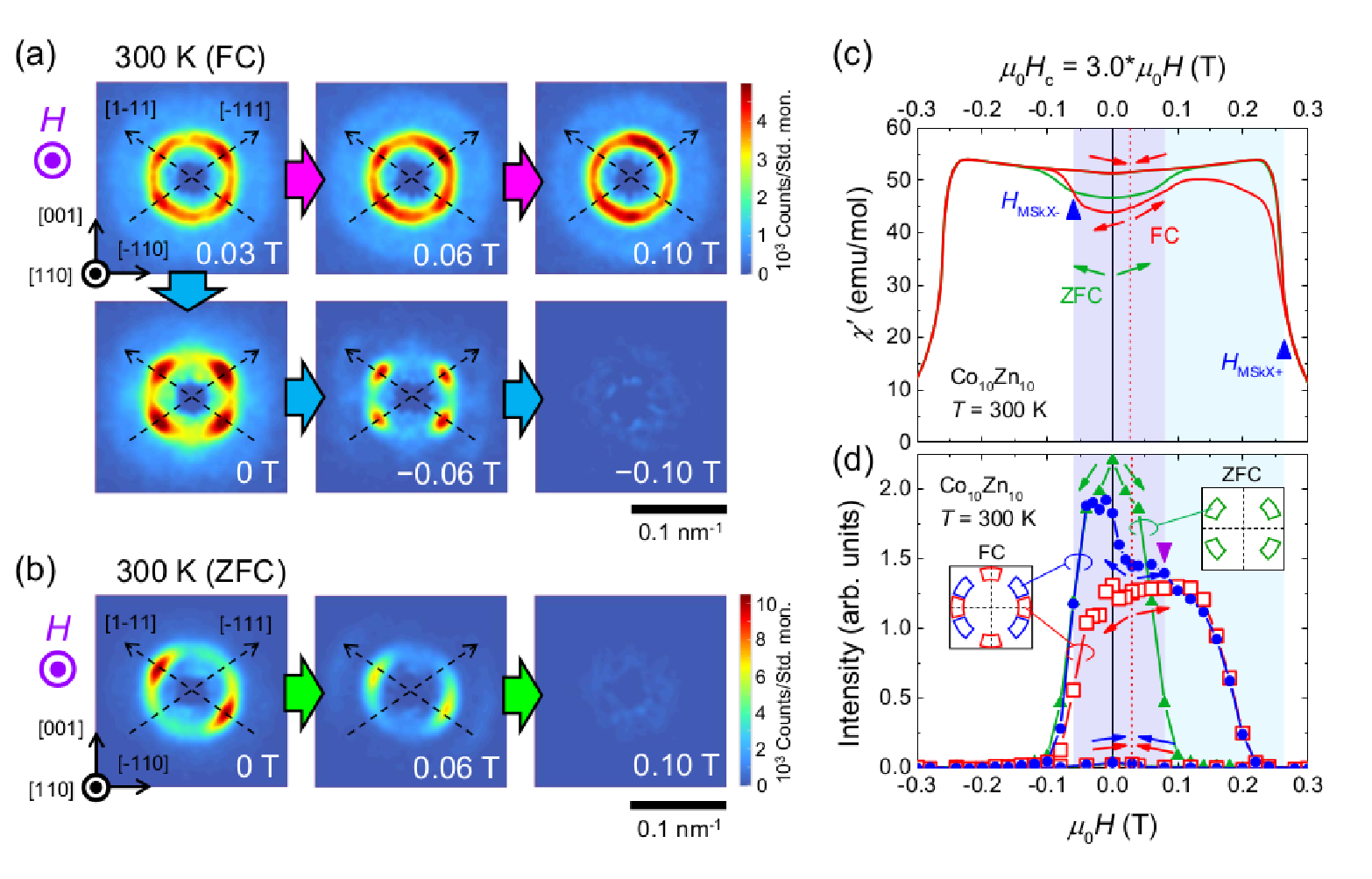}
\end{center}
\end{figure}
%%%%%%%%%%%%%%%%%  FIGS2  %%%%%%%%%%%%%%
\noindent
FIG. S2. Field dependence of metastable SkX state and helical state in \ten\ at 300 K. 
(a) SANS images observed at selected fields during field scans to the positive and negative directions at 300 K after a field cooling (FC) at 0.03 T. The intensity scale of the color plot is fixed through the panels. 
(b) SANS images observed at 0 T, 0.06 T and 0.1 T at 300 K after a zero-field cooling (ZFC). The intensity scale of the color plot is fixed for these panels. 
For panels (a) and (b), the [-111] and [1-11] directions are presented with broken arrows.
(c) Field dependence of the real part of the ac susceptibility ($\chi^\prime$) at 300 K after FC with $\mu_0 H_\mathrm{c}$ = 0.03 T (red line) and ZFC (green line). The blue triangles ($H_\mathrm{MSkX+(-)}$) denote the boundaries of metastable SkX that are plotted in the state diagram in Fig. 2(b) in the main text. The boundary at the positive field side is determined as the field where the hysteresis in $\chi^\prime$ vanishes, and the boundary at the negative field side is determined as the inflection point. 
(d) Field dependence of the SANS intensities at 300 K after the FC at 0.03 T and ZFC. For the FC data, SANS intensity integrated over the azimuthal angle areas at $\phi$ = 55$^\circ$, 125$^\circ$, 235$^\circ$, 305$^\circ$ with the width of $\Delta\phi$ = 30$^\circ$ (blue area in the inset) and at $\phi$ = 0$^\circ$, 90$^\circ$, 180$^\circ$, 270$^\circ$ with the width of $\Delta\phi$ = 30$^\circ$ (red area in the inset) are presented by blue closed circles and red open squares, respectively. 
The purple triangle shows the boundary between the metastable rhombic SkX and the triangular one and is determined as the field where the former intensity (blue circle) becomes higher than the later (red square).
For the ZFC data, SANS intensity integrated over the region at $\phi$ = 55$^\circ$, 125$^\circ$, 235$^\circ$, 305$^\circ$ with the width of $\Delta\phi$ = 30$^\circ$ (green area in the inset) is indicated with green closed triangles. Here, $\phi$ is defined as the clockwise azimuthal angle from the vertical direction. 
In panels (c) and (d), the regions of the metastable rhombic SkX and the triangular one are indicated with purple and light-blue shadings, respectively. 
\\

\newpage

%%%%%%%%%%%%%%%%%  FIGS3  %%%%%%%%%%%%%%
\begin{figure}[htbp]
\begin{center}
\includegraphics[width=16cm]{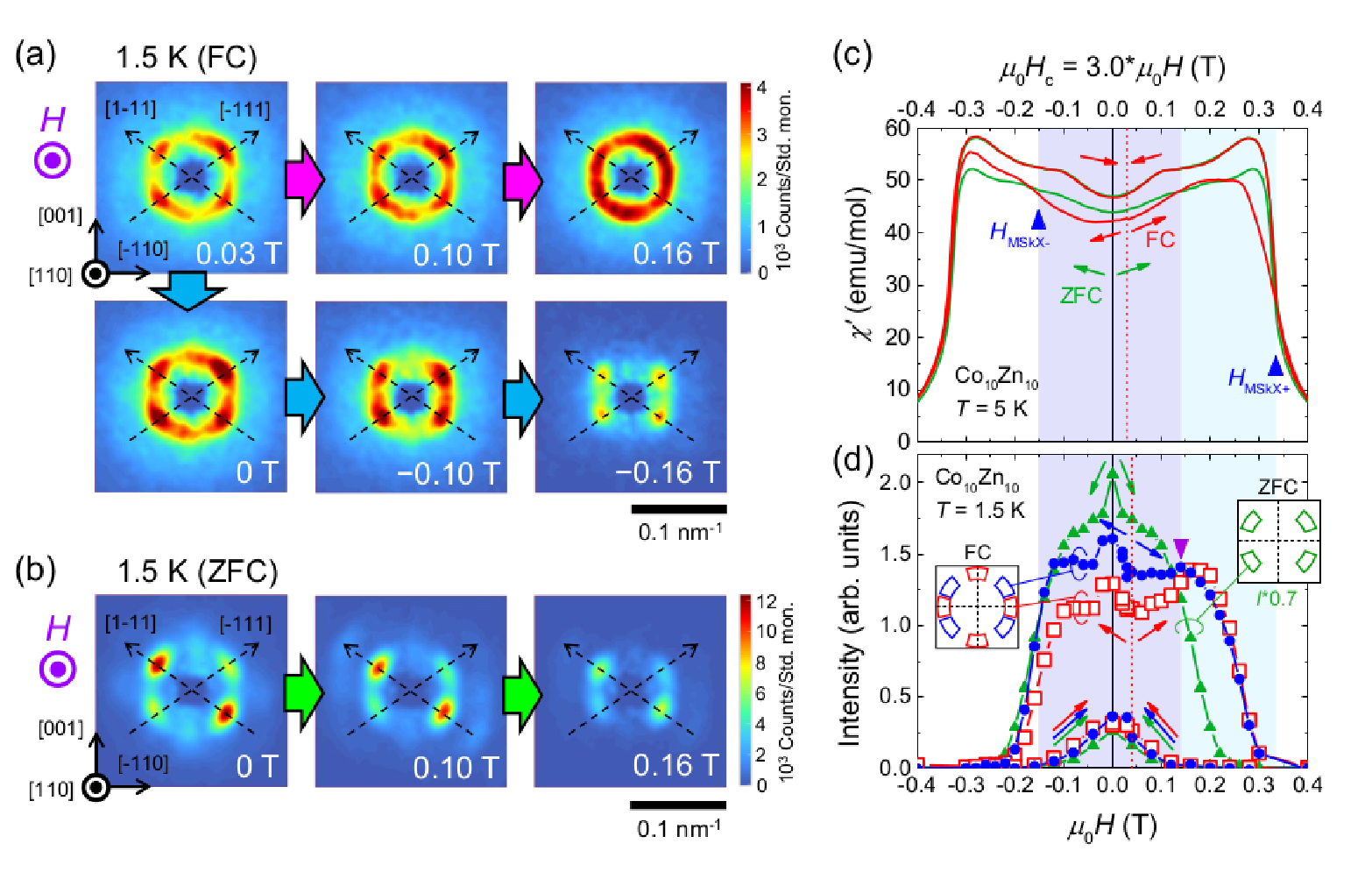}
\end{center}
\end{figure}
%%%%%%%%%%%%%%%%%  FIGS3  %%%%%%%%%%%%%%
\noindent
FIG. S3. Field dependence of metastable SkX state and helical state in \ten\ at 1.5 K. 
(a) SANS images observed at selected fields during field scans to the positive and negative directions at 1.5 K after a FC (0.03 T). The intensity scale of the color plot is fixed through the panels. 
(b) SANS images observed at 0 T, 0.1 T and 0.16 T at 1.5 K after a ZFC. The intensity scale of the color plot is fixed for these panels. 
(c) Field dependence of $\chi^\prime$ at 5 K after the FC with $\mu_0 H_\mathrm{c}$ = 0.03 T (red line) and ZFC (green line). 
The boundaries of metastable SkX ($H_\mathrm{MSkX+(-)}$, blue triangles) are determined similarly as in Fig. S2(c)
(d) Field dependence of the SANS intensities at 1.5 K after the FC (0.03 T) and ZFC, integrated over the azimuthal angle areas defined similarly as in Fig. S2(d).
The boundary between the rhombic SkX and the triangular one (purple triangle) is determined similarly as in Fig. S2(d).
In panels (c) and (d), the regions of the metastable rhombic SkX and the triangular one are indicated with purple and light-blue shadings, respectively.
\\

\newpage

%%%%%%%%%%%%%%%%%  FIGS4  %%%%%%%%%%%%%%
\begin{figure}[htbp]
\begin{center}
\includegraphics[width=14cm]{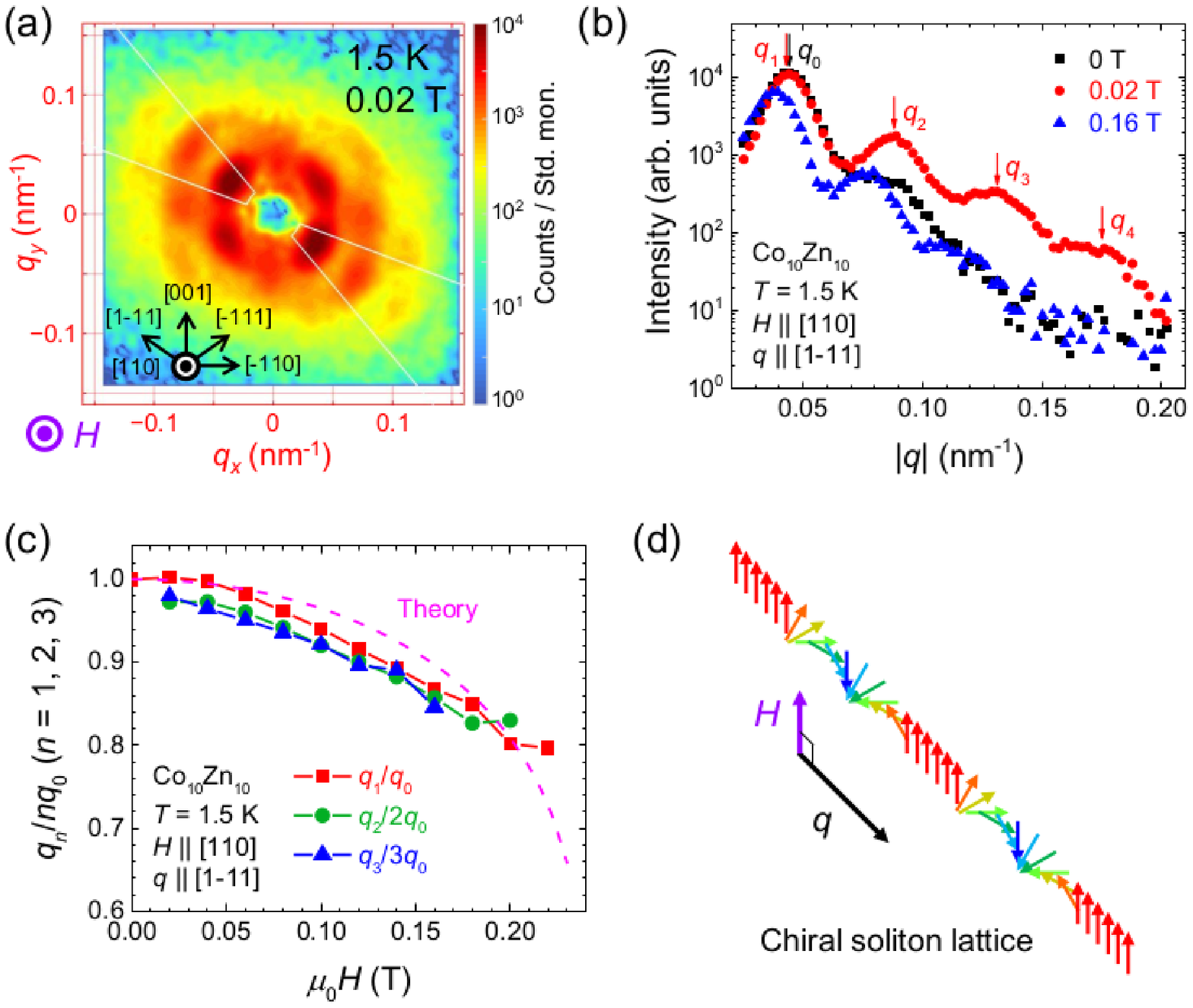}
\end{center}
\end{figure}
%%%%%%%%%%%%%%%%%  FIGS4  %%%%%%%%%%%%%%
\noindent
FIG. S4. Chiral soliton lattice formed under magnetic fields at 1.5 K in \ten. 
(a) SANS image measured at 1.5 K and 0.02 T after ZFC. The intensity of the color plot is displayed in a logarithmic scale.
(b) Radial $|q|$ dependence of SANS intensities (logarithmic scale) measured at 0 T (black), 0.02 T (red) and 0.16 T (blue) at 1.5 K after ZFC. The intensities are integrated over the azimuthal angle area at $\phi$ = 125$^\circ$, 305$^\circ$ with the width of $\Delta\phi$ = 30$^\circ$ [white frames in panel (a)]. The black arrow ($q_0$) indicates the peak position at 0 T. The red arrows ($q_1$, $q_2$, $q_3$ and $q_4$) for 0.02 T indicate the main peak position, second, third and fourth harmonic peak positions, respectively. (c) Normalized $q$ values, $q_1/q_0$, $q_2/2q_0$ and $q_3/3q_0$, are plotted as a function of magnetic field.
Dashed pink line indicates a theoretical curve for the chiral soliton lattice (see text for details). (d) Schematic spin configuration of chiral soliton lattice.
\\

\newpage

%%%%%%%%%%%%%%%%%  FIGS5  %%%%%%%%%%%%%%
\begin{figure}[htbp]
\begin{center}
\includegraphics[width=16cm]{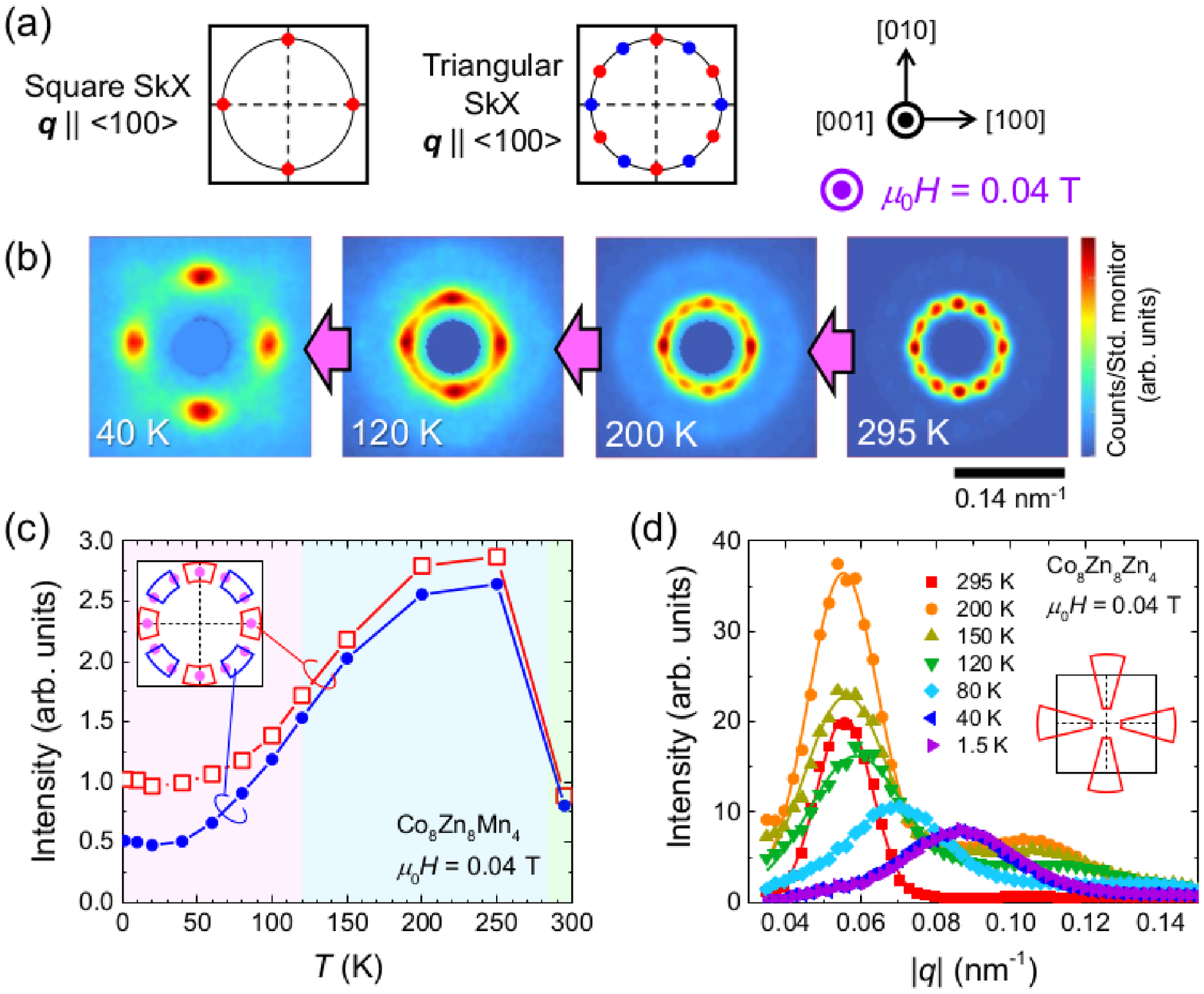}
\end{center}
\end{figure}
%%%%%%%%%%%%%%%%%  FIGS5  %%%%%%%%%%%%%%
\noindent
FIG. S5. Temperature dependence of metastable SkX state in \eight. 
(a) Schematic SANS pattern on the (001) plane, perpendicular to the field, expected for a triangular SkX state (right panel) and for a square SkX state (left panel). The triangular SkX state forms two domains with one of triple-$\vect{q}$ $\parallel$ [100] (6 blue spots) and  $\parallel$ [010] (6 red spots), respectively, resulting in 12 spots at every 30$^\circ$ from $\phi$ = 0$^\circ$. Here, $\phi$ is defined as the clockwise azimuthal angle from the vertical [010] direction. The square SkX state with double-$\vect{q}$ $\parallel$ [100] and [010] shows 4 spots at $\phi$ = 0$^\circ$, 90$^\circ$, 180$^\circ$, 270$^\circ$.
(b) SANS images observed at 295 K, 200 K, 120 K and 40 K on the field cooling (FC) process at 0.04 T (reproduced from our previous work in Ref. \cite{Karube_884}. Copyright 2016, Springer Nature). 
The intensity scale of the color plot varies between each panel. The magnetic field was applied at 295 K after zero-field cooling. 
(c) Temperature dependence of the SANS intensities. Blue closed circles show the SANS intensity integrated over the azimuthal angle area at $\phi$ = 45$^\circ$, 135$^\circ$, 225$^\circ$, 315$^\circ$ with the width of $\Delta\phi$ = 30$^\circ$ (blue area in the inset). Red open squares represent the SANS intensity integrated over the azimuthal angle area at $\phi$ = 0$^\circ$, 90$^\circ$, 180$^\circ$, 270$^\circ$ with the width of $\Delta\phi$ = 30$^\circ$ (red area in the inset). 
The temperature regions of the equilibrium triangular SkX, the metastable triangular SkX and the metastable square SkX are indicated with the light-green, light-blue and pink shadings, respectively.
(d) Radial $|q|$ dependence of SANS intensity, integrated over the azimuthal angle area at $\phi$ = 0$^\circ$, 90$^\circ$, 180$^\circ$, 270$^\circ$ with the width of $\Delta\phi$ = 30$^\circ$ (red area in the inset), at several temperatures on the FC process. 
The data points are fitted to a Gaussian function (solid line).
\\

\newpage

%%%%%%%%%%%%%%%%%  FIGS6  %%%%%%%%%%%%%%
\begin{figure}[htbp]
\begin{center}
\includegraphics[width=14cm]{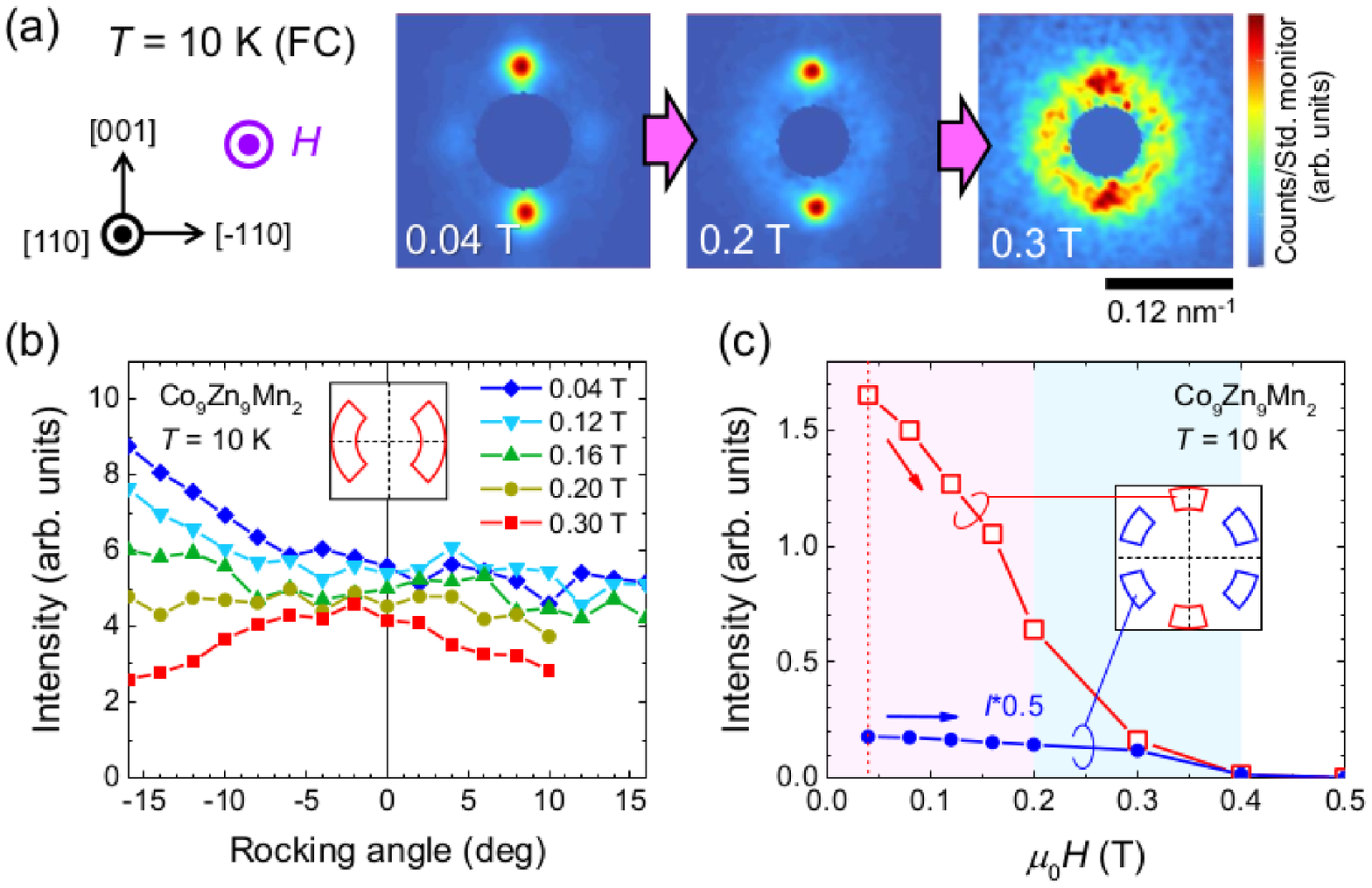}
\end{center}
\end{figure}
%%%%%%%%%%%%%%%%%  FIGS6  %%%%%%%%%%%%%%
\noindent
FIG. S6. Field dependence of metastable SkX state in \nine\ at 10 K. 
(a) SANS images at 0.04 T, 0.2 T and 0.3 T measured at 10 K after a field cooling (FC) at 0.04 T. 
The intensity scale of the color plot varies between each panel. 
(b) Rocking curves at several fields in the field-increasing process after the FC. SANS intensities are integrated over the azimuthal angle area at $\phi$ = 90$^\circ$, 270$^\circ$ with the width of $\Delta\phi$ = 90$^\circ$ (red area in the inset). 
Here, $\phi$ is defined as the clockwise azimuthal angle from the vertical [001] direction. 
(c) Field dependence of the SANS intensities.
Blue closed circles show the SANS intensity integrated over the azimuthal angle area at $\phi$ = 60$^\circ$, 120$^\circ$, 240$^\circ$, 300$^\circ$ with the width of $\Delta\phi$ = 30$^\circ$  (blue area in the inset) and finally divided by 2. 
Red open squares represent the SANS intensity integrated over the azimuthal angle area at $\phi$ = 0$^\circ$, 180$^\circ$ with the width of $\Delta\phi$ = 30$^\circ$ (red area in the inset).
The field regions of the metastable triangular SkX and the metastable square SkX are indicated with the light-blue and pink shadings, respectively.
\\

\newpage